\documentclass[prd,superscriptaddress,nofootinbib,preprintnumbers,amsmath,amssymb,10pt,twocolumn]{revtex4-1}
\usepackage[utf8]{inputenc}
\usepackage{amsfonts}
\usepackage{mathrsfs}
\usepackage{physics}
\usepackage{footnote}
\usepackage{scrextend}
\usepackage{leftidx}
\usepackage{soul}
\usepackage{braket}
 
\usepackage[dvipsnames]{xcolor}
\definecolor{red}{rgb}{0.9, 0,0}
\definecolor{cerulean}{rgb}{0., 0.42,0.9}

\usepackage{epsfig}
\usepackage{graphicx}  
\usepackage{url}
\usepackage{color}
\usepackage{hyperref}
\hypersetup{
     colorlinks   = true,
     citecolor    = red,
	 linkcolor=blue
}
\usepackage{float}
\usepackage[export]{adjustbox}

\usepackage[lofdepth,lotdepth,caption=false,justification=RaggedRight]{subfig}

\renewcommand{\bra}{\langle}
\renewcommand{\ket}{\rangle}

\definecolor{red}{rgb}{0.9, 0,0}
\definecolor{cerulean}{rgb}{0., 0.62,0.9}
\definecolor{navy}{rgb}{0.05, 0.05,0.8}

\graphicspath{ {./figures/} }

\newcommand{\bfe}{{\bf e}}

\newcommand{\bfq}{{\bf q}}

\newcommand{\bfr}{{\bf r}}
\newcommand{\bfl}{\ensuremath{\boldsymbol\ell}}
\newcommand{\bfk}{{\bf k}}

\newcommand{\bfv}{{\bf v}}
\newcommand{\bfu}{{\bf u}}

\newcommand{\code}{\texttt{DarkELF}}

\begin{document}

\title{
Dark matter direct detection from the single phonon to the nuclear recoil regime}
\author{Brian Campbell-Deem}
\affiliation{ Department of Physics, University of California, San Diego, CA 92093, USA }
\author{Simon Knapen}
\affiliation{Theoretical Physics Group, Lawrence Berkeley National Laboratory, Berkeley, CA 94720, USA}
\affiliation{Berkeley Center for Theoretical Physics, Department of Physics, University of California, Berkeley, CA 94720, USA}
\affiliation{CERN, Theoretical Physics Department, Geneva, Switzerland}
\author{Tongyan Lin}
\affiliation{ Department of Physics, University of California, San Diego, CA 92093, USA }
\author{Ethan Villarama}
\affiliation{ Department of Physics, University of California, San Diego, CA 92093, USA }

\date{\today}

\begin{abstract}\noindent In most direct detection experiments, the free nuclear recoil description of dark matter scattering breaks down for masses   $\lesssim$ 100 MeV, or when the recoil energy is comparable to a few times the typical phonon energy. For dark matter lighter than 1 MeV, scattering via excitation of a single phonon dominates and has been computed previously, but for the intermediate mass range or higher detector thresholds, multiphonon processes dominate. We perform the first calculation of the scattering rate via multiphonon production for the entire keV-GeV dark matter mass range, assuming a harmonic crystal target. We provide an analytic description that connects the single phonon, multiphonon, and the nuclear recoil regimes. Our results are implemented in the public package \texttt{DarkELF}.

\end{abstract}

\maketitle

\tableofcontents

\section{Introduction}
\label{sec:intro}

The effort to directly detect dark matter (DM) is entering the sub-GeV mass regime, thanks to experimental innovations which allow for ever lower energy thresholds. For kinematic reasons, this regime is especially challenging for DM which primarily couples to hadronic matter. For a DM mass ($m_\chi$) below 1 GeV, the energy that the DM can deposit in an elastic collision with a nucleus of mass $m_N$ is bounded by 
\begin{equation} \label{eq:simplekinematics}
E_N \leq \frac{2v^2m_\chi^2}{m_N}.
\end{equation}
For $m_\chi\ll m_N$ this is only a small fraction of the total available DM kinetic energy, which can make it very difficult to detect.
This problem can be mitigated to some extent by choosing light element targets such as H~\cite{NEWS-G:2017pxg}, He~\cite{Guo:2013dt,Liu:2016qwd,Hertel:2018aal}, or diamond~\cite{Kurinsky:2019pgb} and by pushing for lower thresholds. Alternatively, one may leverage inelastic processes such as the Migdal effect \cite{Vergados:2004bm,Bernabei:2007jz,Ibe:2017yqa} or bremsstrahlung \cite{Kouvaris:2016afs}.  Inelastic processes occur at substantially lower rate, but are not subject to the constraint in \eqref{eq:simplekinematics} and can also yield signals that are more easily detected than a nuclear recoil, such as electronic excitations,  ionizations or X-rays. Which approach is preferable depends on the characteristics of the detector.

At sufficiently low energy and momentum scales, DM-nucleus scattering is also not subject to \eqref{eq:simplekinematics}  because atom-atom interactions become important. In particular, the relevant excitations in a crystal target are phonons instead of elastic nuclear recoils. For $m_\chi \lesssim$ MeV, the momentum transfer from DM scattering corresponds to wavelengths comparable or larger than the interatomic spacing of a typical target. In this regime, the dominant process will be coherent scattering off multiple atoms, with creation of a single phonon. For crystalline targets with phonon energies as high as $\sim$100 meV, the energy deposited from DM can be well above the naive estimate in \eqref{eq:simplekinematics}. Single phonon excitation has been studied extensively for sub-MeV dark matter, where numerical and analytic calculations by different groups are in good agreement~\cite{Knapen:2017ekk,Griffin:2018bjn,Trickle:2019nya,Cox:2019cod,Griffin:2019mvc,Trickle:2020oki,Griffin:2020lgd,Coskuner:2021qxo}. These calculations have also been extended to diphonon production from sub-MeV dark matter\footnote{Analogous calculations were performed for superfluid He \cite{Schutz:2016tid,Knapen:2016cue,Acanfora:2019con,Caputo:2019cyg,Caputo:2020sys,Baym:2020uos,Matchev:2021fuw}, for which diphonon production is the leading observable process for $m_\chi\lesssim$ 1 MeV.}~\cite{Campbell-Deem:2019hdx} as well as to single phonon production from MeV-GeV dark matter by including Umklapp processes~\cite{Trickle:2019nya,Griffin:2019mvc}. However, so far there has not been a complete description of DM scattering for intermediate energy and momentum transfers, where multiphonon processes are expected to dominate.

\begin{figure}
\centering
\includegraphics[width=0.97\linewidth]{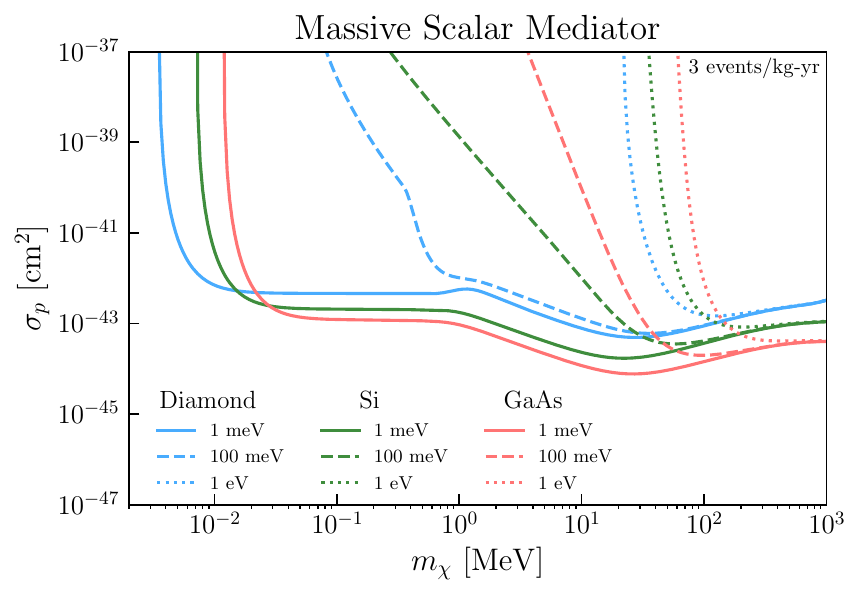} 
\caption{Cross sections needed for 3 events/kg-year for various target materials and threshold energies. A massive hadrophilic mediator is assumed.  
\label{fig:intro_summary}}
\end{figure}

In this work, we develop an analytic treatment of DM scattering that interpolates between the single phonon and nuclear recoil regimes. The relevant approximations are set primarily by the momentum transfer $q$. For single phonon excitations and $q < 2 \pi/a$, where $a$ is typical atomic lattice spacing, we use a long-wavelength approximation  used earlier in the literature \cite{Knapen:2017ekk,Griffin:2018bjn,Cox:2019cod,Campbell-Deem:2019hdx}. For $q > 2 \pi/a$,  we employ the incoherent approximation,  which neglects interference effects between the response of neighboring atoms. This allows us to organize the calculation as a systematic expansion in the number of final state phonons, where each additional phonon comes with a factor of $q/\sqrt{2 m_d \bar \omega_d}$. Here, $m_d$ and $\bar \omega_d$ are the mass and average oscillation frequency of the atom in the position indexed by $d$.  For $q<\sqrt{2 m_d \bar \omega_d}$ it is numerically practical to compute the rate order-by-order in terms of the phonon density of states of the material. For $q \gg \sqrt{2 m_d \bar \omega_d}$, scattering into many phonons dominates and the perturbation series requires increasingly large orders in $q/\sqrt{2 m_d \bar \omega_d}$ to converge. It can however be resummed by making use of the impulse approximation, which in turn smoothly matches onto the free nuclear recoil regime. A similar expansion in number of modes has been performed previously for the integrable toy model that is the harmonic oscillator \cite{Kahn:2020fef}. Here we have generalized the approach to a harmonic \emph{crystal}, analogous to the procedure followed in \cite{Knapen:2020aky} and \cite{Berghaus:2021wrp}, in calculations of the Migdal effect and X-ray backgrounds, respectively.   Fig.~\ref{fig:intro_summary} illustrates our results from applying these approximations. All of our calculations are implemented as part of the \texttt{DarkELF} public code~\cite{Knapen:2021bwg}.\footnote{\url{https://github.com/tongylin/DarkELF}} 

The remainder of this paper is organized as follows: In Sec.~\ref{sec:structurefactor} we introduce the dynamic structure factor, which captures the material-dependence of the DM scattering cross section, and motivate the incoherent approximation for the structure factor. In Sec.~\ref{sec:processes}, we describe our analytic approximations in detail across the different regimes in energy and momentum transfer. We perform checks on our use of the incoherent approximation by comparing with previous calculations for single-phonon production and analytic calculations for diphonon production. Our results for GaAs are discussed in detail in Sec.~\ref{sec:results} and we conclude in Sec.~\ref{sec:discuss}. Appendix~\ref{sec:two_phonon_appendix} contains the formulas for diphonon production and Appendix~\ref{app:impulse} provides details on the impulse approximation. The  implementation in \texttt{DarkELF} is documented in Appendix~\ref{app:results_darkelf}. We further provide numerical results for Ge, Si and diamond in Appendix~\ref{app:extraresults}.

\section{Dynamic structure factor}
\label{sec:structurefactor}
Our starting point will be a general potential for spin-independent DM-nucleus interactions, although the formalism below could also be applied to spin-dependent interactions.  For a DM particle of mass $m_\chi$ incident on a crystal with $N$ unit cells and $\mathfrak{n}$ ions per unit cell, the potential in Fourier space is given by
\begin{equation}
    \label{eq:potential}
    \Tilde{\mathcal{V}}(\bfq) =  \frac{2 \pi b_p }{\mu_\chi}\Tilde{F}(\mathbf{q}) \sum\displaylimits_{\bfl}^N \sum\displaylimits_{d=1}^{\mathfrak{n}}  f_{\bfl d} e^{i \mathbf{q} \cdot \mathbf{r}_{\bfl d}}.
\end{equation}
Here, we sum over the $N$ unit cells, labeled by lattice vectors $\bfl$, and atoms within the unit cell, labeled with the index $d$, such that all atoms in the crystal with positions $\mathbf{r}_{\bfl d}$ are summed over. The DM-proton scattering length $b_p$ is  defined by the DM-proton scattering cross section $\sigma_p \equiv 4 \pi b_p^2$ at some reference momentum, and $\mu_\chi$ is the DM-proton reduced mass. We first consider a general coupling strength $f_{\bfl d}$ of the nucleus labeled by $\bfl,d$ relative to that of a single proton. $f_{\bfl d}$ is specified for various interactions in Section \ref{sec:results}, such as nucleon number for scalar mediators and the effective electric charge for scattering via a dark photon mediator. In the latter case $f_{\bfl d}$ is $\bfq$ dependent when accounting for screening effects. 

We consider two form factors in \eqref{eq:potential} representing limiting cases of interactions: scattering via a heavy mediator, where $\Tilde{F}(\mathbf{q})=1$; and scattering via a massless mediator, where $\Tilde{F}(\mathbf{q}) = q_0^2/q^2$ with a model-dependent reference momentum $q_0$. 

Collecting the overall factor $2\pi b_p \Tilde{F}(\mathbf{q})/\mu_\chi$, we define the differential cross section as
\begin{equation}
    \label{eq:cross-sec-eqn}
    \frac{d \sigma}{d^3 \bfq \, d\omega} = \frac{b_p^2}{\mu_\chi^2} \frac{1}{v} \frac{\Omega_c}{2\pi} |\Tilde{F}(\bfq)|^2 S(\bfq, \omega) \, \delta\left(\omega - \omega_\bfq \right)
\end{equation}
where $v$ is the initial velocity of the dark matter (incident on a target at rest), $\Omega_c = V/N$ is the volume of the unit cell in the crystal, and $\omega_\bfq = \bfq \cdot \bfv - q^2 / 2m_\chi$ is the kinematic constraint on the momentum and energy transfers to the crystal $\bfq$ and $\omega$. We have in turn also defined the dynamic structure factor 
\begin{equation}
    \label{eq:structure1}
    S(\mathbf{q}, \omega) \equiv \frac{2\pi}{V} \sum\displaylimits_f \left| \sum\displaylimits_{\bfl}^N \sum\displaylimits_{d=1}^{\mathfrak{n}}  \mel{\Phi_f}{f_{\bfl d} e^{i \mathbf{q} \cdot \mathbf{r}_{\ell d}}}{0} \right|^2 \delta\left( E_f - \omega \right).
\end{equation}
Note that the convention for $S(\bfq, \omega)$ varies across the literature; here we use the convention that gives a similar $S(\bfq, \omega)$ definition for both phonon interactions and DM-electron interactions~\cite{Trickle:2019nya,Kahn:2021ttr}. We also assume the system is initially in its ground state $\left| 0 \right\ket$ prior to the collision, corresponding to a zero temperature system. We sum over final states with energies $E_f$, such that each term represents the probability to excite the final state $\left| \Phi_f \right\ket $.

\subsection{Coherent and incoherent structure factors}

For a given crystal there are many possible configurations of interaction strengths $f_{\bfl d}$ which may vary even for different samples of the same material, e.g.~the exact distribution of spins or isotopes in the material for spin-dependent\footnote{For spin-dependent interactions, $f_{\bfl d}$ is an operator rather than a parameter, but otherwise the analysis proceeds analogously.} or mass-dependent interactions, respectively. This can be accounted for by averaging over a large collection of target samples. With a large number of nuclei in the crystal, we expect the exact distribution of interaction strengths in a given sample to be inconsequential relative to the result averaged over many samples. We can keep track of fluctuations away from the average configuration by splitting the scattering rate into a coherent and incoherent contribution, as explained below.

We follow the procedure of Refs.~\cite{Schober2014,Squires1996} and first re-express \eqref{eq:structure1} by expanding the square and Fourier transforming the $\delta$-function, giving 
\begin{align}  
    \label{eq:structure_unaveraged}
    S(\bfq,\omega) &= \sum_{\bfl, \, \bfl^\prime}^{N} \sum_{d, \, d^\prime}^{\mathfrak{n}} f_{\bfl d} f_{\bfl^\prime d^\prime}^{*} \,  \mathcal{C}_{\bfl^\prime d^\prime \bfl d}
\end{align}
where $ \mathcal{C}_{\bfl^\prime d^\prime \bfl d}$ is the time-dependent two-point function:
\begin{align}
    \mathcal{C}_{\bfl^\prime d^\prime \bfl d} \equiv& \, \frac{1}{V} \int\displaylimits_{-\infty}^{\infty} \! dt\, \sum_f \mel{0}{ e^{-i \mathbf{q} \cdot \bfr_{\bfl^\prime d^\prime}(0)}}{\Phi_f}\nonumber\\
    & \qquad \qquad \quad \times\mel{\Phi_f}{ e^{i \mathbf{q} \cdot \mathbf{r}_{\bfl d}(t)}}{0}  e^{-i \omega t}\nonumber\\
    \equiv& \, \frac{1}{V} \int\displaylimits_{-\infty}^{\infty} \! dt\, \langle e^{-i \mathbf{q} \cdot \bfr_{\bfl^\prime d^\prime}(0)} e^{i \mathbf{q} \cdot \mathbf{r}_{\bfl d}(t)} \rangle e^{-i \omega t}.\label{eq:correlation}
\end{align}
In the second line we used the completeness of the basis of states. It will also be advantageous to define a shorthand notation for the \textit{auto-correlation function} for an atom with itself as
\begin{align}\label{eq:autocorrelationdef}
    \mathcal{C}_{\bfl d} &\equiv \, \mathcal{C}_{\bfl d \bfl d} \nonumber\\
    &\equiv \, \frac{1}{V} \int\displaylimits_{-\infty}^{\infty} \! dt\, \langle e^{-i \mathbf{q} \cdot \bfr_{\bfl d}(0)} e^{i \mathbf{q} \cdot \mathbf{r}_{\bfl d}(t)} \rangle e^{-i \omega t}.
\end{align}

We assume that the $f_{\bfl d}$ are random throughout the crystal. Under this assumption, the average of $f_{\bfl d} f^*_{\bfl' d'}$ over target configurations,  $\overline{f_{d} f^*_{d'}}$,  must be independent of the lattice sites $\bfl, \bfl'$. 
Making this replacement in \eqref{eq:structure_unaveraged} gives
\begin{align}
    \label{eq:structure2}
    S(\bfq,\omega) &= \sum_{\bfl, \, \bfl^\prime}^{N} \sum_{d, \, d^\prime}^{\mathfrak{n}} \overline{f_{d} f_{d^\prime}^{*}} \,  \mathcal{C}_{\bfl^\prime d^\prime \bfl d}
\end{align}
where the averages may be written as
\begin{align*}
   d &\neq d^\prime: & \overline{f_{d} f_{d^\prime}^{*}} &= \overline{f_{d}} \, \overline{f^{*}_{d^\prime}}, \\
    d &= d^\prime: & \overline{f_{d} f_{d^\prime}^{*}} &= \overline{f_{d}^2} .
\end{align*}
For the $d\neq d'$ case we assumed that the expectation values of the $f_d$ for different atoms in the unit cell are uncorrelated.
This allows one to split the structure factor into two contributions:
\begin{align}
    S(\bfq, \omega) =& \sum_{\bfl \neq \bfl^\prime}^{N} \sum_{d \neq d^\prime}^{\mathfrak{n}} \overline{f_{d}} \, \overline{f^{*}_{d^\prime}} \, \mathcal{C}_{\bfl^\prime d^\prime \bfl d}
    +  \sum_{\bfl}^{N} \sum_{d}^{\mathfrak{n}}  \overline{f_{d}^2}  \, \mathcal{C}_{\bfl d}\\
    =&   \sum_{\bfl,  \bfl^\prime}^{N} \sum_{d, d^\prime}^{\mathfrak{n}}\overline{f_{d}} \, \overline{f^{*}_{ d^\prime}}  \, \mathcal{C}_{\bfl^\prime d^\prime \bfl d} +  \sum_{\bfl}^{N}  \sum_{d}^{\mathfrak{n}} \left(\overline{f_{d}^2} - (\overline{f_{d}})^2 \right)  \, \mathcal{C}_{\bfl d}\label{eq:cohdef}\\
    \equiv&\,  S^\mathrm{(coh)}(\bfq, \omega) +   S^\mathrm{(inc)}(\bfq, \omega)  \label{eq:inc-coh}
\end{align}
where the second line is obtained by adding and subtracting the term proportional to $(\overline{f_{d}})^2$ and regrouping.
The first and second term in \eqref{eq:inc-coh} are usually referred to as the \textit{coherent} and \textit{incoherent} structure factors in the neutron scattering literature.

The coherent structure factor relays the scattering rate if the interaction strengths of all atoms in equivalent lattice sites are equal to a common value $\overline{f_d}$. For example, one can consider low energy, spin-independent neutron scattering in a very pure crystal with only a single isotope per atom type. This implies $f_{d}=f_{\bfl d}=A_d$, with $A_d$ the atomic mass number, such that the incoherent contribution in \eqref{eq:inc-coh} vanishes exactly. The sum in \eqref{eq:cohdef} then crucially includes position correlators between differing nuclei, which capture the interference between different lattice sites. In practice, this interference leads to a coherence condition, which demands that the momentum in the scattering process must be conserved up to a reciprocal lattice vector. In particular, the 0th order term in a low $\bfq$ expansion of \eqref{eq:correlation} corresponds to Bragg diffraction.

The incoherent structure factor on the other hand accounts for the statistical variations in interaction strengths between different scattering centers in the lattice. The second sum in \eqref{eq:cohdef} contains no cross terms and thus does not include interference between different lattice sites. There is therefore no corresponding coherence condition and the incoherent structure factor does not enforce momentum conservation.\footnote{An alternative but equivalent point of view is that for coherent scattering, translation symmetry is broken up to a shift symmetry, since all unit cells are identical. For incoherent scattering the scattering centers are treated as independent and translation invariance is therefore fully broken, resulting in the complete loss of momentum conservation.} 

For most earlier DM direct detection calculations the focus has been on spin-independent scattering in high purity crystals with little isotopic variation. In this scenario, we take the single isotope approximation $\overline{f_{d}^2} - (\overline{f_{d}})^2 = 0$, implying that only the coherent scattering contributes. For spin-dependent dark matter scattering, the average will be the quantum expectation value of the spin operator, resulting in  $\overline{f_{d}^2} \neq (\overline{f_{d}})^2$. We therefore expect the incoherent piece in \eqref{eq:inc-coh} to be important in this case. In this paper we focus exclusively on spin-independent scattering in the single isotope limit and the  corresponding coherent structure factors. The coherent structure factors are however more difficult to evaluate, due to the conservation of crystal momentum that is built into \eqref{eq:correlation}. This results in increasingly complicated phase space integrals for multiphonon processes~\cite{Campbell-Deem:2019hdx}. For our purposes, the utility of studying the incoherent structure factor will be that the auto-correlation function can be used to obtain a reasonable and more manageable approximation of the coherent structure factor at sufficiently high momenta. Our results can also be extended to the case of spin-dependent scattering, but we leave this for future work. 

Before venturing further into this approximation and its validity, we must first develop the structure factors into a form which lends itself to a direct calculation. In order to evaluate the structure factors in \mbox{\eqref{eq:structure1}--\eqref{eq:structure2}}, the position vector of each atom may be decomposed in terms of the equilibrium lattice positions and displacement vectors, $\bfr_{\bfl d} = \bfl + \bfr_d^0 + \bfu_{\bfl d}$. Here $\bfr_d^0$ is the equilibrium location of atom $d$ relative to the origin of the primitive cell and $\bfu_{\bfl d}$ is the displacement relative to that equilibrium.
Following this decomposition, we quantize the displacement vector in the harmonic approximation with a phonon mode expansion 
\begin{align}
    \mathbf{u}_{\bfl d}(t) = \sum\displaylimits_{\nu}^{3 \mathfrak{n}} \sum\displaylimits_{\mathbf{k}}& \frac{1}{\sqrt{2N m_d \omega_{\nu, \mathbf{k}}}} \Big( {\bf e}_{\nu, d, \mathbf{k}} \hat{a}_{\nu, \mathbf{k}} e^{i \mathbf{k} \cdot (\bfl + \mathbf{r}_d^0 ) - i \omega_{\nu, \bfk} t} \nonumber \\
    &+ {\bf e}^*_{\nu, d, \mathbf{k}} \hat{a}^{\dagger}_{\nu, \mathbf{k}} e^{-i \mathbf{k} \cdot (\bfl + \mathbf{r}_d^0 ) + i \omega_{\nu, \bfk} t} \Big)
\label{eq:displacement}
\end{align}
The index $\nu$ denotes the phonon branches, of which there are $3 \mathfrak{n}$, and $\bfk$ labels the phonon momentum in the first Brillouin Zone (BZ). The $\hat{a}^{\dagger}_{\nu, \mathbf{k}}$ and $\hat{a}_{\nu, \mathbf{k}}$ are the creation and annihilation operators for the phonons, $\omega_{\nu, \mathbf{k}}$ is the energy of the phonon, $\mathbf{e}_{\nu, d, \mathbf{k}}$ is the phonon eigenvector for atom $d$ normalized within a unit cell, $\sum_d \mathbf{e}_{\nu, d, \mathbf{k}}^*\cdot\mathbf{e}_{\mu, d, \mathbf{k'}} = \delta_{\mu \nu} \delta_{\bfk, \bfk'}$, and $m_d$ is the mass of atom $d$. 

The structure factor in \eqref{eq:structure2} can then be explicitly evaluated by applying \eqref{eq:displacement} to \eqref{eq:correlation}. For a pure single isotopic crystal with $\overline{ f^2_d} = (\overline{ f_d})^2$, this is given by~\cite{Campbell-Deem:2019hdx}
\begin{widetext}
\begin{equation}
    \label{eq:coherent_previous}
    S^\mathrm{(coh)}(\bfq, \omega) = \frac{2\pi}{V} \sum\displaylimits_{f} \left| \sum\displaylimits_{\bfl}^{N} \sum\displaylimits_{d}^{\mathfrak{n}} \overline{f_d} \,  e^{-W_d(\bfq)} \mathcal{M}_{\bfl d} \right|^2 \delta \left( E_f - \omega \right)
\end{equation}
where
\begin{equation}
\label{eq:melement}
     \mathcal{M}_{\bfl d} \equiv  e^{i\bfq \cdot (\bfl + \bfr^{0}_{d})} \left\bra \Phi_f \right| \exp \left[ i \sum\displaylimits_{\bfk, \nu}\frac{\bfq \cdot \bfe^*_{\nu, \bfk, d}}{\sqrt{2 N m_d \omega_{\nu, \bfk}}} \hat{a}^{\dagger}_{\nu, \bfk} e^{-i \bfk \cdot (\bfl + \bfr^0_d)}\right] \left| 0 \right\ket
\end{equation}
\end{widetext}
is the matrix element for scattering into the final state of the crystal denoted by $f$. The Debye-Waller factor $e^{-W_d(\bfq)}$ is given in terms of the function $W_d(\mathbf{q}) \equiv  \frac{1}{2} \langle \left( \mathbf{q} \cdot \mathbf{u}_{\bfl d} (0) \right)^2 \rangle$. We may Taylor expand the inner exponential in powers of $\bfq$ where the $n$th term can excite a final state consisting of $n$ phonons. The phonon eigenvectors and energies may be obtained numerically using Density Functional Theory (DFT) (see e.g.~\cite{phonopy}); using these, the leading single phonon structure factor has been calculated~\cite{Griffin:2018bjn,Griffin:2019mvc,Coskuner:2021qxo}. These DFT-based calculations quickly become cumbersome, however, and have not yet been performed for generic $n$-phonon terms.
Analytic calculations may be performed more easily, and have been carried out for the single- and two-phonon terms~\cite{Campbell-Deem:2019hdx}, but are only tractable when assuming an isotropic crystal and that $|\bfq|$ is small relative to the size of the first Brillouin zone. Such analytic calculations likewise lack scalability for higher order phonon terms.

In summary, since the direct evaluation of \eqref{eq:coherent_previous} is very tedious and not always possible, we will rely instead on an approximate form of $S^\mathrm{(coh)}(\bfq, \omega)$, bypassing the need to deal with \eqref{eq:coherent_previous}. This is described in the next section.

\subsection{Incoherent approximation\label{sec:incoherentapprox}}

The incoherent approximation amounts to dropping the cross terms in ($\bfl\neq\bfl'$ or $d\neq d'$) from the sum in \eqref{eq:cohdef}, thus neglecting the interference between non-identical atoms. In other words, one approximates the coherent structure factor by
\begin{align}
        \label{eq:inc-approx}
       S^\text{(coh)}(\bfq, \omega) &\approx \sum_{\bfl}^{N}  \sum_{d}^{\mathfrak{n}} (\overline{f_{d}})^2 \mathcal{C}_{\bfl d}.
\end{align}
The incoherent structure factor remains unchanged, and the total structure factor is then given by $S^\text{(tot)}(\bfq, \omega) \approx \sum_{\bfl}^{N}  \sum_{d}^{\mathfrak{n}} \overline{f_{d}^2} \mathcal{C}_{\bfl d}$. In this work we will focus only on pure crystals with a single isotope for each type of atom, so that the total structure factor can be computed with \eqref{eq:inc-approx}.
The incoherent approximation is expected to be a good approximation when the momentum transfer is larger than $2 \pi/a$ with $a$ the inter-particle spacing. Then the phase factors associated with the interference terms are expected to add up to a small correction compared to the $\bfl=\bfl', d=d'$ terms in the sum. For an argument justifying \eqref{eq:inc-approx} we refer to \cite{Schober2014,PhysRev.82.392}.

For momentum transfers within the first Brillouin zone, single phonon scattering always dominates the inclusive scattering rate. It is however possible that the detector threshold is such that single phonon processes cannot be accessed but the double or multiphonon processes can. In this case the incoherent approximation cannot  a priori be taken for granted. We nevertheless use it, but verify the results against our earlier two-phonon calculations \cite{Campbell-Deem:2019hdx} whenever possible (Sec.~\ref{sec:twophonon}), finding satisfactory agreement. The accuracy of the calculations in this part of phase space is however less well understood and further work is needed.

To evaluate the auto-correlation function, we first replace the atomic positions $\bfr_{\bfl d}$ in \eqref{eq:inc-coh} with their displacement operator decomposition, noting that the $\bfl + \bfr^0_d$ constant cancels, amounting to a simple substitution of $\bfr_{\bfl d} \rightarrow \bfu_{\bfl d}$:
\begin{equation}
   \mathcal{C}_{\bfl d} = \frac{1}{V} \int\displaylimits_{-\infty}^{\infty} \! dt\, \langle e^{-i \mathbf{q} \cdot \bfu_{\bfl d}(0)} e^{i \mathbf{q} \cdot \mathbf{u}_{\bfl d}(t)} \rangle e^{-i \omega t}
\end{equation}
The expectation value may be rewritten with an application of the Baker–Campbell–Hausdorff formula, Bloch's identity $\langle e^{\hat A} \rangle=e^{\frac{1}{2} \expval{\hat{A}^2}}$, and some matrix algebra~\cite{Griffin:2018bjn} giving:
\begin{equation}
    \label{eqn:incohmigdal}
    \mathcal{C}_{\bfl d} = \frac{1}{V} \int\displaylimits_{-\infty}^{\infty} \! dt\, e^{-2 W_d(\mathbf{q})} e^{\langle \mathbf{q} \cdot \mathbf{u}_{\bfl d}(0) \, \mathbf{q} \cdot \mathbf{u}_{\bfl d} (t) \rangle} e^{-i\omega t}.
\end{equation}
When we deployed Bloch's identity, we implicitly used the harmonic approximation, by only considering displacement operators of the form in \eqref{eq:displacement}. 

\begin{figure}
\includegraphics[width=0.48\textwidth]{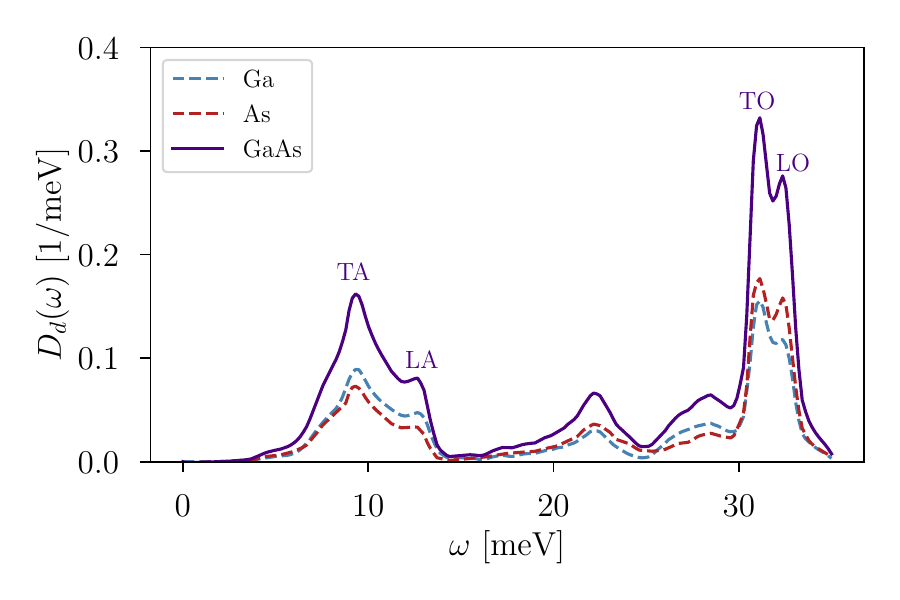}
\caption{Partial and total density of states for GaAs \cite{Jain2013}. Labels indicate the regions in which a particular phonon branch dominates.\label{fig:densityofstates}}
\end{figure}

The correlator $\langle \mathbf{q} \cdot \mathbf{u}_{\bfl d}(0) \, \mathbf{q} \cdot \mathbf{u}_{\bfl d} (t) \rangle$ may be evaluated with the form of the displacement operator in \eqref{eq:displacement}, wherein the $\bfl$ dependence cancels. This gives 
 \begin{align}
\langle \bfq \cdot \bfu_{d}(0) \, \bfq \cdot \bfu_{d}(t) \rangle &= \sum_{\nu} \sum_{\bfk} \frac{ \left| \bfq \cdot \bfe_{\nu, \bfk, d} \right|^2 }{2 N  m_d \omega_{\nu, \bfk}}  e^{i \omega_{\nu,\bfk} t}\\
\intertext{which can be simplified further by averaging over the direction of momentum vector $\bfq$ }
\langle \bfq \cdot \bfu_{d}(0) \, \bfq \cdot \bfu_{d}(t) \rangle &\approx\frac{q^2}{3}\sum_{\nu} \sum_{\bfk} \frac{ |\bfe_{\nu, \bfk, d} |^2 }{2 N  m_d \omega_{\nu, \bfk}}  e^{i \omega_{\nu,\bfk} t}\\
&= \frac{ q^2}{2 m_d}\int\displaylimits_{-\infty}^{+\infty}\!\! d\omega'\frac{D_d(\omega') }{\omega'}  e^{i \omega' t}\label{eq:correldensitystates}
 \end{align}
where we defined the \emph{partial density of states} for each atom in the primitive cell as
\begin{equation}
D_d(\omega)\equiv \frac{1}{3N }\sum_{\nu} \sum_{\bfk} |\bfe_{\nu, \bfk, d} |^2 \delta(\omega-\omega_{\nu,\bfk}).
\end{equation}
The partial density of states was normalized to satisfy $\int_{-\infty}^{+\infty}\!\!d\omega D_d(\omega)=1$. This can be shown by using the eigenvector completeness condition, which imposes $\sum_\nu {e}_{\nu, \bfk, d, i}^{*} e_{\nu, \bfk, d, j} = \delta_{ij}$ for fixed $\bfk, d$, where $i, \,j$ are spatial indices.
In addition, the total density of states of the material is defined by
\begin{equation}
D(\omega)\equiv \sum_d D_d(\omega)=\frac{1}{3N}\sum_{\nu} \sum_{\bfk}  \delta(\omega-\omega_{\nu,\bfk}),
\end{equation}
which satisfies $\int_{-\infty}^{+\infty}\!\!d\omega D(\omega)=\mathfrak{n}$ with $\mathfrak{n}$ the number of atoms in the unit cell.\footnote{In the literature, the density of states is also sometimes normalized to $3n_a$, where $n_a$ is the atomic density.} In materials such as Ge, Si, or GaAs all atoms in the primitive cell have the same or similar mass and as such contribute roughly equally to the density of states, see Fig.~\ref{fig:densityofstates}. One could therefore approximate $D_d(\omega)\approx D(\omega)/\mathfrak{n}$ in \eqref{eq:correldensitystates} for these materials. We however choose to keep track of the partial density of states, to keep the calculations as general as possible.

For mono-atomic lattices, the density of states can be extracted directly from neutron scattering data through the incoherent structure factor. This is not always possible for multi-atomic lattices, since the scattering is only sensitive to the combination $\sum_d |\overline{ f_d}|^2 D_d(\omega)/m_d $. To infer the individual $D_d(\omega)$ as well as $D(\omega)$, one therefore needs a set of scattering techniques which allows one to effectively vary the $\overline{f_d}$. This is not available for all materials, and it is therefore often most convenient to extract the $D_d(\omega)$ from DFT calculations. A comprehensive library of results has been made available by the materials project \cite{Jain2013}. 
 
Returning now to the calculation of the autocorrelation function, we can expand the exponential term in \eqref{eqn:incohmigdal} using the form of the correlator in \eqref{eq:correldensitystates}. This yields an explicit representation of $C_{\ell d}$ as an expansion in number of phonons $n$ being excited:
 \begin{align}
    \label{eqn:n-incoherent}
       \mathcal{C}_{\bfl d} =& \frac{2 \pi}{V}  e^{-2 W_d(\mathbf{q})}\sum_n \frac{1}{n!}  \left(\frac{q^2}{2 m_d} \right)^n  \nonumber \\ \times &\left(\prod\displaylimits_{i=1}^n \int d \omega_{i} \frac{D_d(\omega_i)}{\omega_i} \right) \delta \left(\sum_j \omega_j - \omega \right)
 \end{align}
where the delta function arises from the time integral $\frac{1}{2\pi} \int dt\, e^{i(\sum \omega_i)t} e^{-i \omega t}$ and ensures energy conservation. Here, by using \eqref{eq:correldensitystates}, the Debye-Waller function takes the form of 
\begin{equation}
\label{eq:DWapprox}
    W_d(\bfq) = \frac{q^2}{4 m_d} \int d\omega^\prime \frac{D_d(\omega^\prime)}{\omega^\prime}.
\end{equation}
Thus, in comparison to the difficulties discussed surrounding \eqref{eq:coherent_previous}, inputting this form of the correlator into \eqref{eq:inc-approx} gives an analytic approximation for all phonon terms in the appropriate regime of validity.
 
In this paper, we utilize the incoherent approximation to calculate the contributions from higher-order phonon terms to an arbitrary degree in a simple and fast manner.  This allow us to make rate predictions for the entire relevant mass range, going from the low-mass ($m_\chi \gtrsim $ keV) single phonon regime to the high-mass ($m_\chi \gtrsim $ 50 MeV) nuclear recoil regime.

 
 \section{Processes \label{sec:processes}}
 
Using the autocorrelation function, \eqref{eqn:n-incoherent}, we can estimate the scale at which a generic $n$-phonon term starts becoming a relevant contribution to scattering. To organize the multiphonon expansion, it is useful to define an average phonon energy 
\begin{align}\label{eq:omegaddeff}
    \bar \omega_d \equiv \int d \omega^\prime \omega^\prime D_d(\omega^\prime).
\end{align}
While $\bar \omega_d$ technically depends on the atom $d$, this just gives an $\mathcal{O}(1)$ dependence in the phonon scale. Since $n! \propto n^n$ at large $n$, we see that the $n$th term of the series \eqref{eqn:n-incoherent} will roughly begin giving an $\mathcal{O}(1)$ contribution when
\begin{align}
    \label{eq:typical-n}
    \frac{q^2}{2 m_d \bar \omega_d} \sim n.
\end{align}
This means that for a given $q$ (or consequently, $m_\chi$) one can determine the dominant scattering processes. When $\frac{q^2}{2 m_d \bar \omega_d} \lesssim 1$, single phonon excitations will be the primary channel; for $m_d \sim 30$ GeV and $\bar \omega_d\sim 30$ meV, this corresponds to $q \lesssim 30$ keV. Conversely, when $\frac{q^2}{2 m_d \bar \omega_d} \gg 1$, phonons are no longer a suitable description and the scattering is instead well modeled by the recoil of a single nucleus. This transition occurs roughly at $q \gtrsim 2 \sqrt{2 m_d \bar\omega_d}$.  In between these two extremes, we have $n \sim$ few, indicating multiphonon excitations as the primary process. The precise nature of the dominant process for a given $m_\chi$ will vary based on the mediator mass and experimental threshold.

In this section, we describe analytic approaches for characterizing the structure factor in crystal targets, broken into subsections corresponding to the previously mentioned processes. Secs.~\ref{sec:singlephonon} and \ref{sec:twophonon} deal with single phonon and two phonon excitations. Here we can also compare calculations of the full structure factor with the incoherent approximation. Sec.~\ref{sec:multiphonon} deals with many phonon excitations, and Sec.~\ref{sec:impulse_approx} describes the impulse approximation, which gives a good approximation to the structure factor for momenta approaching the nuclear recoil limit. For all numerical results in this section, we will assume a coupling to nucleons (replacing the generic average interaction strength $\overline{f_d}$ with the nucleon number $A_d$) for both massive and massless mediators, and take a GaAs target as a typical example of a simple cubic crystal of interest. 

\subsection{Single phonon production}
\label{sec:singlephonon}
If the unit cell contains at least two atoms, there are two types of phonons that can be produced: \textit{acoustic} and \textit{optical} phonons. As discussed in Sec.~\ref{sec:structurefactor}, DFT-based calculations for both single acoustic and single optical phonon excitations have been performed across a large dark matter mass range ($\sim$keV to GeV)~\cite{Griffin:2018bjn,Griffin:2019mvc,Coskuner:2021qxo}. Meanwhile analytic calculations so far have been limited $q\lesssim1$ keV, which corresponds to $m_\chi \lesssim$ MeV~\cite{Knapen:2017ekk,Campbell-Deem:2019hdx}.  Although the DFT-based calculations span the entire mass range of interest and can provide information such as directional dependence, the numerics are more intensive; the phonon band structure, eigenvectors  and structure factors must be calculated from first principles for each material. For high $q$, the sum over the reciprocal lattice must also be accounted for \cite{Trickle:2019nya,Griffin:2020lgd}.
Here we extend the analytic calculations to the high $q$ regime by using the incoherent approximation. The comparison with the DFT results of  \cite{Griffin:2018bjn} will also serve as a validation of the incoherent approximation.

To organize the calculations, it is useful to define a momentum scale ($q_{\mathrm{BZ}}$) which approximately reflects the size of the first Brillouin zone. We take $q_\mathrm{BZ} = \frac{2 \pi}{a}\approx 2$ keV, where $a$ is the lattice constant. We first review the single phonon response for $q<q_\mathrm{BZ}$. 
In this regime, we compute the structure factors in the isotropic approximation and in the limit $q\ll q_\mathrm{BZ}$. For this purpose we assume linear dispersions $\omega=c_s q$ for the longitudinal acoustic (LA) and transverse accoustic (TA) modes, with $c_s$ replaced by $c_\mathrm{LA}$ and $c_\mathrm{TA}$ for the longitudinal and transverse sound speeds, respectively. The optical modes are assumed to have flat (constant) dispersions for the longitudinal optical (LO) and transverse optical (TO) phonon energies $\omega_\mathrm{LO}$ and $\omega_\mathrm{TO}$. The sound speeds and optical phonon energies are taken to be their long-wavelength values $(q=0)$. We will refer to this set of assumptions as the \emph{long-wavelength approximation}.

The matrix element is given by the leading non-trivial term in the small $q$ expansion of \eqref{eq:melement}. The only relevant contributions for $q\ll q_{\rm BZ}$ are those of the single LA and LO phonons.
We approximate the long-wavelength acoustic eigenvectors as 
\begin{equation}\label{eq:accoustic_eigenvectors}
\bfe_{\mathrm{LA},\bfk,d}\approx \frac{\sqrt{A_d} }{\sqrt{ \sum_{d'} A_{d'}}} \hat{\bfk};
\end{equation} note that this form is valid for generic crystal targets and not limited to GaAs. For the LO phonon, we use the following eigenvectors, which are only valid for diatomic lattices~\cite{Campbell-Deem:2019hdx}
\begin{align}
\label{eq:optical_eigenvectors} \bfe_{\mathrm{LO},\bfk, 1} \approx& \frac{\sqrt{A_2}}{\sqrt{A_1 + A_2}} \, \hat{\bfk}, \\ \bfe_{\mathrm{LO},\bfk, 2} \approx& - \frac{\sqrt{A_1}}{\sqrt{A_1 + A_2}} e^{-i  \bfk \cdot \bfr^0_2} \, \hat{\bfk}
\end{align}
where the first atom is taken to be at the origin of the primitive cell, and the second atom is taken to be at the coordinate \mbox{$\bfr^0_2 = (a/4, \, a/4, \, a/4)$} for GaAs.  The acoustic and optical transverse eigenvectors are orthogonal to these, but do not contribute to the scattering into a single phonon. With these approximations and taking $\overline{ f_d} = A_d$, the analytic expressions for the single phonon contributions to the structure factor are~\cite{Campbell-Deem:2019hdx}
\begin{align}
    &S_{n=1, {\rm LA}}(q, \omega) \approx \frac{2\pi}{\Omega_c} \frac{\left( \sum\displaylimits_{d'} A_{d'} \right) q^2}{2 m_p \omega_{\mathrm{LA}, q}} \delta (\omega - \omega_{\mathrm{LA}, q} ) \Theta (\omega_\mathrm{LO} - \omega)
   \label{eq:single-ph-aco-analytic}
    \\
    &S_{n=1, {\rm LO}}(q, \omega) \approx \frac{2\pi}{\Omega_c} \frac{q^4 a^2}{32 \omega_{\mathrm{LO}}} \frac{A_1 A_2}{m_p (A_1 + A_2) } \delta ( \omega - \omega_{\mathrm{LO}})
    \label{eq:single-ph-opt-analytic}
    \\
     &S^{(q < q_{\rm BZ})}_{n=1}(q, \omega) =    S_{n=1, {\rm LA}}(q, \omega) +    S_{n=1, {\rm LO}}(q, \omega)    \label{eq:singlephononsum}
\end{align}
with $\Omega_c$ the volume of the primitive cell. Here we have introduced a cut-off of $\omega = \omega_\mathrm{LO}$ to the longitudinal acoustic branch to avoid overestimating the scattering rate with the LA mode near the edge of the Brillouin zone.
The $q^4$ scaling and appearance of the lattice constant $a$ in the optical structure factor comes from averaging over angles with the eigenvectors, giving $(\bfq \cdot \bfr^0_2 )^2 \approx q^2 a^2 / 16$~\cite{Cox:2019cod}.

For dark matter with a standard velocity dispersion $v \sim 10^{-3}$, the typical momentum transfer begins to fall outside the first Brillouin zone for $m_\chi \gtrsim$ 1 MeV. Physically, this corresponds to the wavelength becoming smaller than the interatomic spacing, and the long-wave length formulas from \eqref{eq:accoustic_eigenvectors} to \eqref{eq:single-ph-opt-analytic} are no longer valid. We can however utilize the incoherent approximation in \eqref{eq:inc-approx} and \eqref{eqn:n-incoherent}, which yields
\begin{align}
    \label{eq:single-ph-inc-approx}
    S^{(q > q_\mathrm{BZ})}_{n=1}(q, \omega) \approx \frac{2\pi}{\Omega_c} \sum\displaylimits_d^{\mathfrak{n}}  &e^{-2 W_d(q)} (\overline{f_{d}})^2 \, \frac{q^2}{2 m_d} \frac{D_d(\omega)}{\omega}.
\end{align}
The forms of the structure factor are qualitatively quite different in the two $q$ regimes. In the coherent regime $q < q_\mathrm{BZ}$, summing over the response of multiple atoms with constructive interference leads to a resonant response in \eqref{eq:singlephononsum}.
The impact of the interference is greatly reduced for $q > q_\mathrm{BZ}$, such that the incoherent approximation becomes a viable description. 

\begin{figure*}
\centering
\subfloat[Comparison of the integrated single phonon structure factor for GaAs. The left panel shows the structure factor integrated over $\omega = 1-27$ meV for acoustic phonon branches only and the right panel has $\omega = 27-40$ meV for optical phonon branches only. The dashed line shows the DFT result, averaged over all $\bfq$ directions, while the solid line shows our analytic approximation based on joining \eqref{eq:singlephononsum} (valid for  $q<q_\mathrm{BZ}$) with the incoherent approximation \eqref{eq:single-ph-inc-approx} (valid for $q>q_\mathrm{BZ}$).
]{\label{fig:singlephonon_top}
\includegraphics[width=0.47\linewidth]{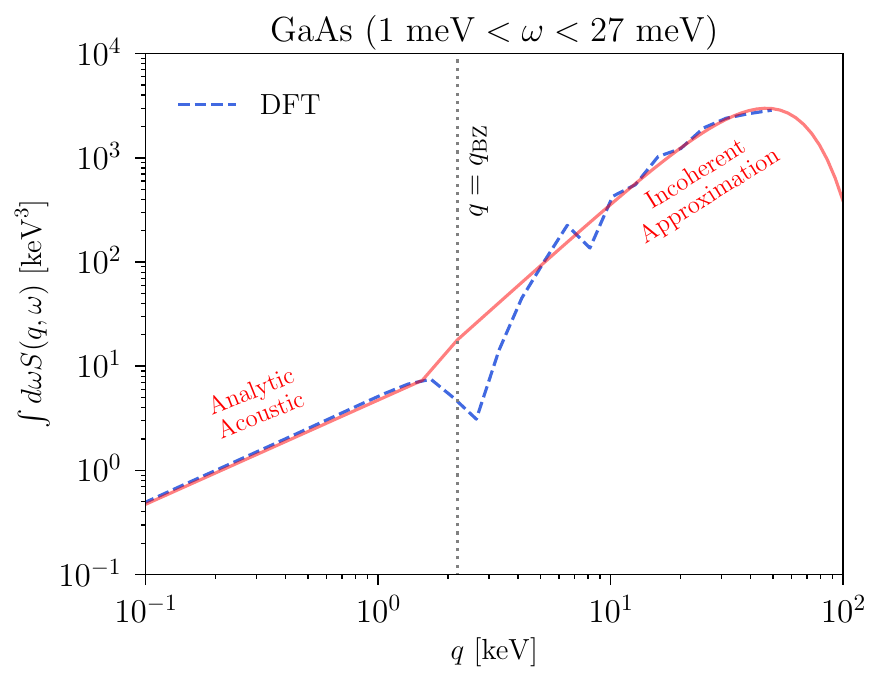} \hfill \includegraphics[width=0.47\linewidth]{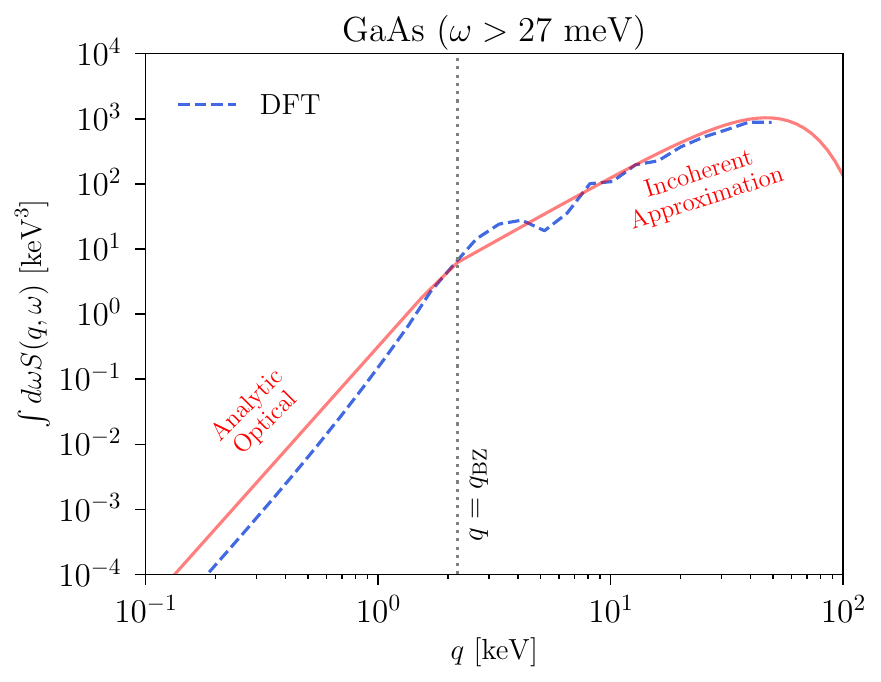}
}
\newline
\subfloat[Cross sections giving a rate of 3 events/kg-year, assuming $\overline{f_d} = A_d$. The rate is computed using our analytic single phonon structure factor approximation (solid) or with DFT calculations (dashed). We find that the analytic approach agrees with the DFT calculations within an $\mathcal{O}(1)$ factor. ]{\label{fig:singlephonon_bottom}
\includegraphics[width=0.47\linewidth]{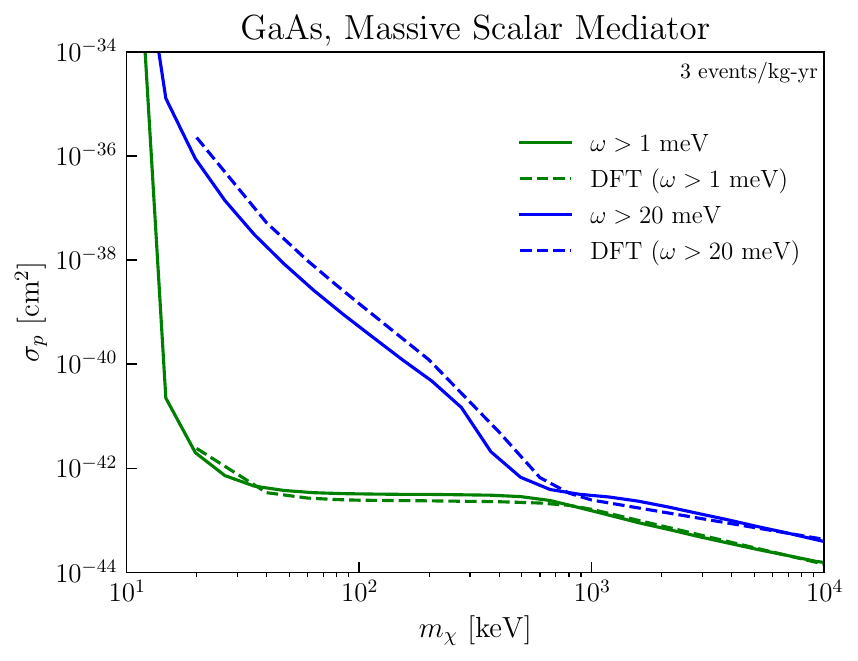} 
\includegraphics[width=0.47\linewidth]{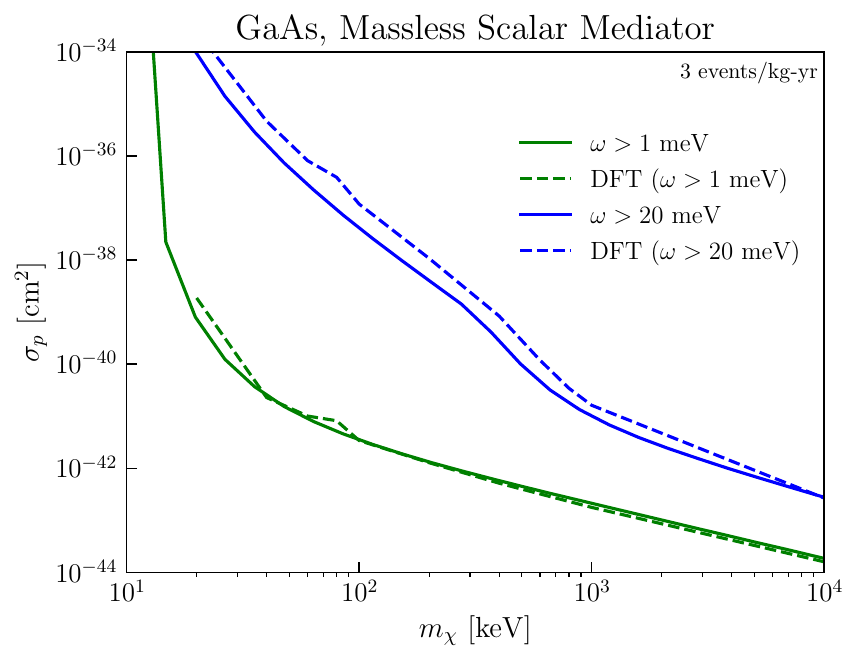}} 
\caption{\textbf{Single phonon production.}}
\label{fig:singlephonon}
\end{figure*}

While the sharp transition in the structure factor is an artifact of our approximations, \eqref{eq:singlephononsum}-\eqref{eq:single-ph-inc-approx} can accurately describe the integrated structure factor above or below $q_{\rm BZ}$.
Fig.~\ref{fig:singlephonon} compares our combined analytic single phonon description with numerical DFT calculations. For the DFT result we follow \cite{Griffin:2018bjn}, computing the dynamical matrix and phonon dispersions with respectively VASP~\cite{PhysRevB.54.11169} and phonopy~\cite{phonopy} (see also~\cite{Griffin:2019mvc}), and take the angular average of $S(\bfq,\omega)$ over all $\bfq$ directions for comparison with the isotropic approximation. The top  panels show the structure factors in \eqref{eq:singlephononsum} as a function of $q$, integrated over $\omega$. The top left panel shows $S(q,\omega)$ integrated over $\omega \in [ 1\, {\rm meV}, 27\, {\rm meV} ]$ to select the acoustic phonon branches only and the top right panel shows the integral over $\omega \in [ 27\, {\rm meV}, 40\, {\rm meV} ]$ for optical phonon branches. The analytic approximations are in good agreement with the DFT result  in their respective regimes of validity. For $q<q_\mathrm{BZ}$, integrating \eqref{eq:singlephononsum} leads to respectively $\sim q$ and $\sim q^4$ scaling, while the incoherent approximation in \eqref{eq:single-ph-inc-approx} always scales as $\sim q^2$. As discussed above, the $\omega$-dependence of the analytic structure factors is quite different in the two regimes, with the coherent structure factor giving a resonant response around the single-phonon dispersion while the incoherent approximation is continuous in $\omega$. However, the integrated result matches the full DFT calculation of the coherent structure factor well, indicating that the analytic approach will be useful in calculating integrated quantities such as rates. Furthermore, the analytic approach provides physical insight into the change in the $q$-scaling of the structure factor in Fig.~\ref{fig:singlephonon_top}.

The plots in Fig.~\ref{fig:singlephonon_bottom} show single phonon integrated rates for both massive and massless scalar mediators. For the massless mediator, scattering into the acoustic phonon specifically favors small $q$ due to the $\propto q^{-4}$ contribution of the mediator form factor. The analytic result of \eqref{eq:single-ph-aco-analytic} therefore applies across the entire DM mass range, as the large $q$ contributions are negligible. For all other cases the structure factor scales with a positive power of $q$ so that large $q$ contributions are the most important. We therefore see a change in slope of the $\sigma_p$ reach around $m_\chi \sim$ MeV, when $q \gtrsim q_\mathrm{BZ}$ becomes kinematically accessible. These features are captured by the $q>q_\mathrm{BZ}$ analytic description from the incoherent approximation, and again agree with the DFT results.

\subsection{Two-phonon production ($q<q_{\rm BZ}$)}
\label{sec:twophonon}

We next turn to the use and accuracy of the incoherent approximation for two-phonon production, in particular for $q< q_{\rm BZ}$. Single phonon production always dominates in this regime if above threshold \cite{Campbell-Deem:2019hdx}. It is however expected that there will be a phase in the experimental program for which the energy threshold will still be too high to access single optical and accoustic phonons, such that the formally subleading double phonon production can be relevant. 

While the incoherent approximation is expected to be the least accurate for $q<q_{\rm BZ}$, it is still useful to compare it with existing analytical results for the structure factor. The analytic results are obtained in the long-wavelength approximation, as defined in Sec.~\ref{sec:singlephonon}. In this limit, the Wilson coefficients of the self-interaction
operators for the acoustic modes can be extracted from the measured or calculated elasticity parameters. With these assumptions, one can explicitly evaluate \eqref{eq:coherent_previous} to second order in $q/\sqrt{m_d \omega}$~\cite{Campbell-Deem:2019hdx}.

In this work, we will extend the long-wavelength calculations to all final states (see Appendix~\ref{sec:two_phonon_appendix}) and compare them with the incoherent approximation. For this purpose we extrapolate the results of Ref.~\cite{Campbell-Deem:2019hdx} to higher $q$ values and make a number of additional assumptions to model the self-interactions of the optical modes, thus giving the complete structure factor.
For these reasons the calculations in this section should however be considered only a toy model of a GaAs-like crystal. 
We will show below that for this toy model and in the limit of small momentum transfer, the incoherent and long-wavelength approximations give qualitatively similar DM scattering rates.

From Ref.~\cite{Campbell-Deem:2019hdx}, the two-phonon structure factor can be written as 
\begin{equation}
S(\bfq, \omega) = S^{\mathrm{(harm)}} (\bfq, \omega) + S^{\mathrm{(anh)}} (\bfq, \omega)
\end{equation}
in the long-wavelength limit. The first term is the structure factor in the harmonic limit (also referred to as the contact piece in \cite{Campbell-Deem:2019hdx}), where anharmonic corrections to the atomic potentials are neglected. It can be obtained by expanding \eqref{eq:melement} to second order, and evaluated analytically in the long-wavelength limit. The second term contains contributions to the structure factor from anharmonic interactions. In order to evaluate this, one needs to include a phonon self-interaction Hamiltonian in computing \eqref{eq:melement}, as described in detail in \cite{Campbell-Deem:2019hdx}. The interactions of acoustic phonons are based on an effective three-phonon Hamiltonian valid in the long-wavelength limit, but to obtain a more complete picture we include a highly approximate three-phonon Hamiltonian for interactions involving optical phonons.  These calculations are summarized in Appendix~\ref{sec:two_phonon_appendix}.

\begin{figure*}
\centering
\includegraphics[width=0.85\linewidth]{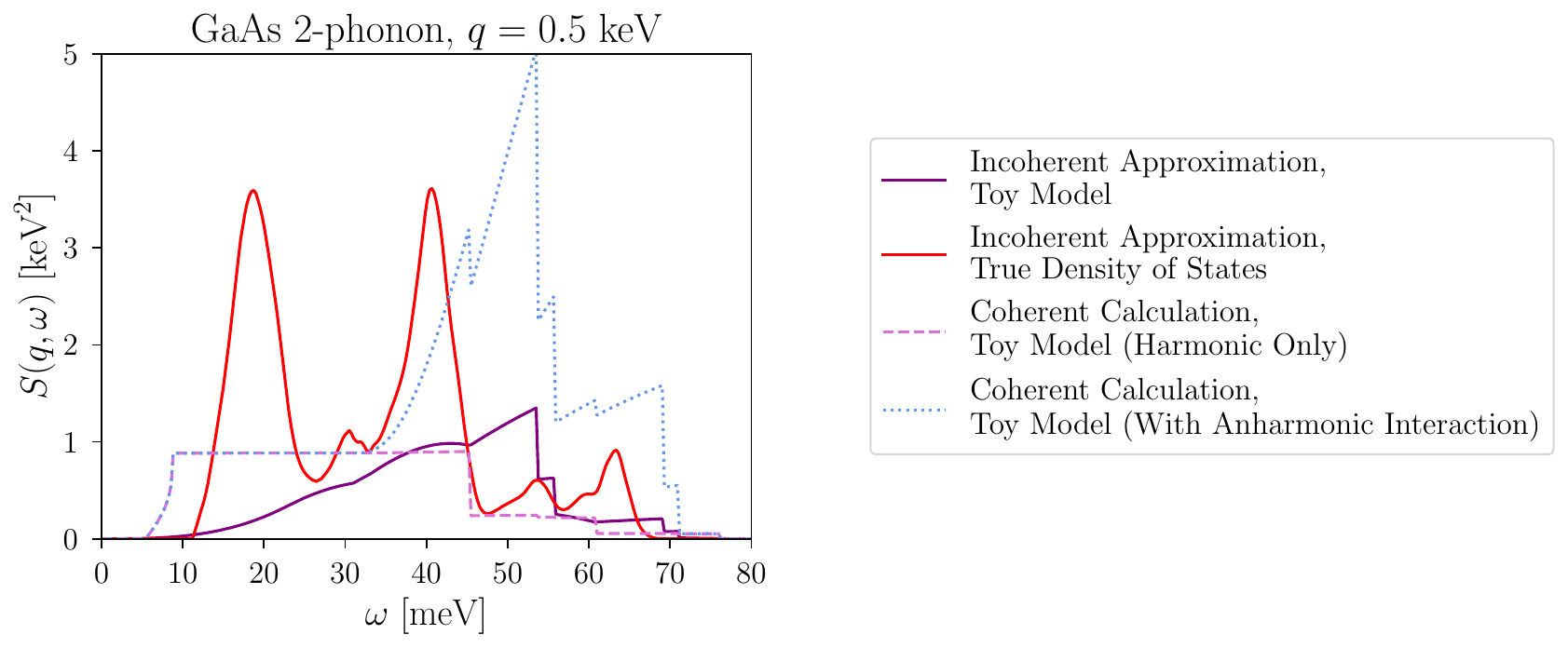} 
\includegraphics[width=0.9\linewidth]{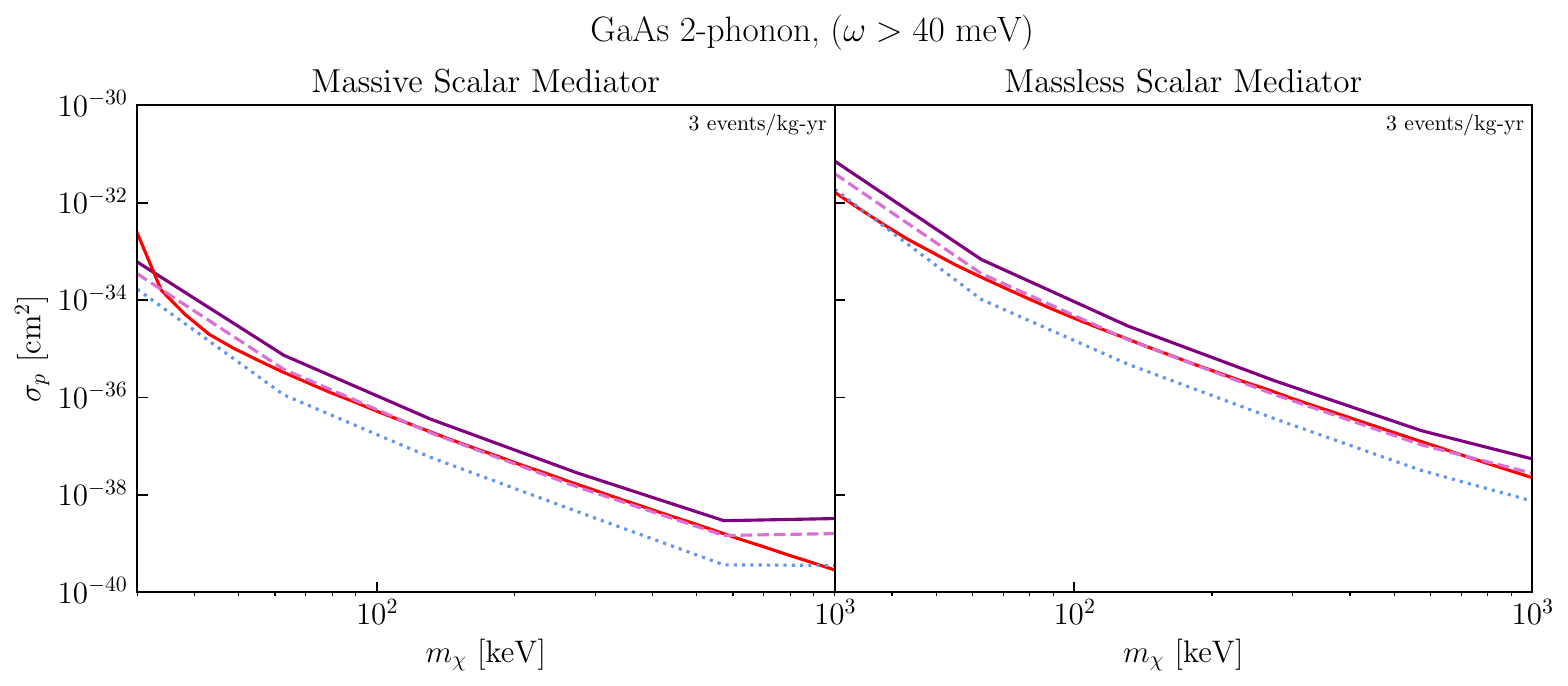}
 \caption{{\bf Two phonon production.}  \emph {Top}: Comparison of the two-phonon structure factor calculated with various approximations, where the toy model assumes the long-wavelength approximation. Optical-optical channels give a $\delta$-function and are not plotted. \emph {Bottom}: Cross sections for producing two phonons at a rate of 3 events/kg-year using the same approximations as above. We restrict the mass range to $m_\chi \lesssim $ 1 MeV so that typical $q$ values are below $q_\mathrm{BZ}$, where our long-wavelength approximations are valid. The energy threshold is taken to be $40$ meV, above the single phonon energies.}
 \label{fig:twophonon}
\end{figure*}

To perform the most meaningful comparison between the incoherent and long-wavelength approximations, we assume the following Debye model for the partial density of states for a diatomic crystal

\begin{align}
\label{eq:GaAs_toy_DoS}
\nonumber
D_{1,2}(\omega) = & \, \frac{1}{q_\mathrm{BZ}^3}\frac{1}{A_1+A_2} \\
& \Bigg(A_{1,2}\frac{\omega^2}{c_\mathrm{LA}^3} \Theta(c_\mathrm{LA}q_\mathrm{BZ} - \omega)\Theta(\omega)\nonumber \\
\nonumber
& + A_{1,2}\frac{2 \omega^2}{c_\mathrm{TA}^3} \Theta(c_\mathrm{TA}q_\mathrm{BZ} - \omega)\Theta(\omega)
\\
&  + A_{2,1}\frac{q_\mathrm{BZ}^3}{3} \delta(\omega - \omega_\mathrm{LO}) \nonumber\\
&+ A_{2,1}\frac{2q_\mathrm{BZ}^3}{3} \delta(\omega - \omega_\mathrm{TO})  \Bigg).
\end{align}
which is derived from the long-wavelength approximation as described in Sec.~\ref{sec:singlephonon}.\footnote{Here the maximum momentum of the modes is determined by requiring that the sum over all modes is equal to the total number of degrees of freedom. For GaAs and in the isotropic approximation, the exact momentum cutoff is about 2\% different from $q_{\rm BZ} = 2\pi/a$. This error is negligible compared to the uncertainties on the other assumptions made in this section.}
The explicit structure factor from using this toy density of states in \eqref{eqn:n-incoherent} is given in Appendix~\ref{sec:two_phonon_appendix}, which for simplicity we evaluate with $A_1 = A_2$ for GaAs.

The top panel of Fig.~\ref{fig:twophonon} compares the calculations of the two-phonon structure factor in the incoherent and long-wavelength approximations. For the incoherent approximation, we show the result with the toy density of states in \eqref{eq:GaAs_toy_DoS} as well as with the true density of states from Fig.~\ref{fig:densityofstates}.
The dashed line shows the harmonic limit, meaning that $S^{(\rm anh)}$ is neglected. This is the case that is most directly comparable to the incoherent approximation, which assumes the harmonic mode expansion in \eqref{eq:displacement}. For the dotted line, the leading phonon self-interactions were included.

In the harmonic limit, all modes scale as $\sim q^4$ except for optical-acoustic final state, which scales as $\sim q^6$. The incoherent approximation naturally misses these more subtle destructive interference effects, but still captures the correct $q^4$ scaling for most of the modes. We see in Fig.~\ref{fig:twophonon} that the incoherent approximation is within a factor of $\sim 5$ of the long-wavelength approximation for all $\omega>\omega_{\rm LO}$, for both the toy model and true density of states. The difference at smaller $\omega$ is not experimentally relevant, as the single phonon rate will completely dominate in this region. There are also delta-function terms from the optical-optical branches which do not appear in the plot; their contributions to the overall scattering rate are comparable for the incoherent and long-wavelength approximations as well. See Appendix~\ref{sec:two_phonon_appendix} for details. 
These terms dominate the scattering rate at higher energies, and overall we see in Fig.~\ref{fig:twophonon} that the incoherent approximation reproduces the structure factor in the harmonic limit to within a factor of few.

When anharmonic interactions are included, the difference becomes larger and the incoherent approximation may under-predict the rate by up to an order of magnitude in our estimate. However, as discussed above, the anharmonic Hamiltonian used is itself also only valid at the order of magnitude level, particularly for optical modes.  We expect that our approach can model the rate in this regime at the order-of-magnitude level, but a proper DFT calculation is needed for it to be rigorously validated.

Finally, we show in the bottom panel of Fig.~\ref{fig:twophonon} a comparison of the cross sections corresponding to a rate of 3 events/kg year, with the different approximations for the two-phonon structure factor. We assume $\omega > 40$ meV, since for lower thresholds the rate is dominated by single-phonon production~\cite{Campbell-Deem:2019hdx}. We emphasize that here we are only illustrating that the incoherent approximation is within a factor of few of the full structure factor, as long as the same assumptions are made for the phonon dispersion relations. Therefore, we restrict our comparison to $m_\chi < $ MeV such that we can restrict to $q < q_\mathrm{BZ}$. The incoherent approximation underestimates the rate by a factor of few in the harmonic limit, and up to an order of magnitude when anharmonic interactions are included. Using the true density of states slightly improves the agreement. Though this comparison only applies to a limited $q$ range, our result suggests that the incoherent approximation should give a reasonable, order-of-magnitude estimate for multiphonon production even at low $q$. We expect this uncertainty to decrease for larger $q$ where the incoherent approximation is most justified, and in particular we will see that the incoherent approximation reproduces the expected rate in the free nuclear recoil limit, as discussed in the next sections.


\subsection{Multiphonon production}
\label{sec:multiphonon}

\begin{figure*}
\centering
\subfloat[The first ten phonon structure factors in the incoherent approximation for GaAs, plotted for various fixed $q$.  At sufficiently large $q>\sqrt{2 m_d \bar \omega_d}$, the total structure factor converges to the impulse approximation (IA, dashed line). In the right panel, there is a slight difference between the peak of the true structure factor and the impulse approximation. This can be accounted for in the impulse approximation by including higher orders in the steepest descent expansion~\cite{Liang:2022xbu}. ]{\label{fig:impulse-structures_top} 
\includegraphics[width=0.95\linewidth]{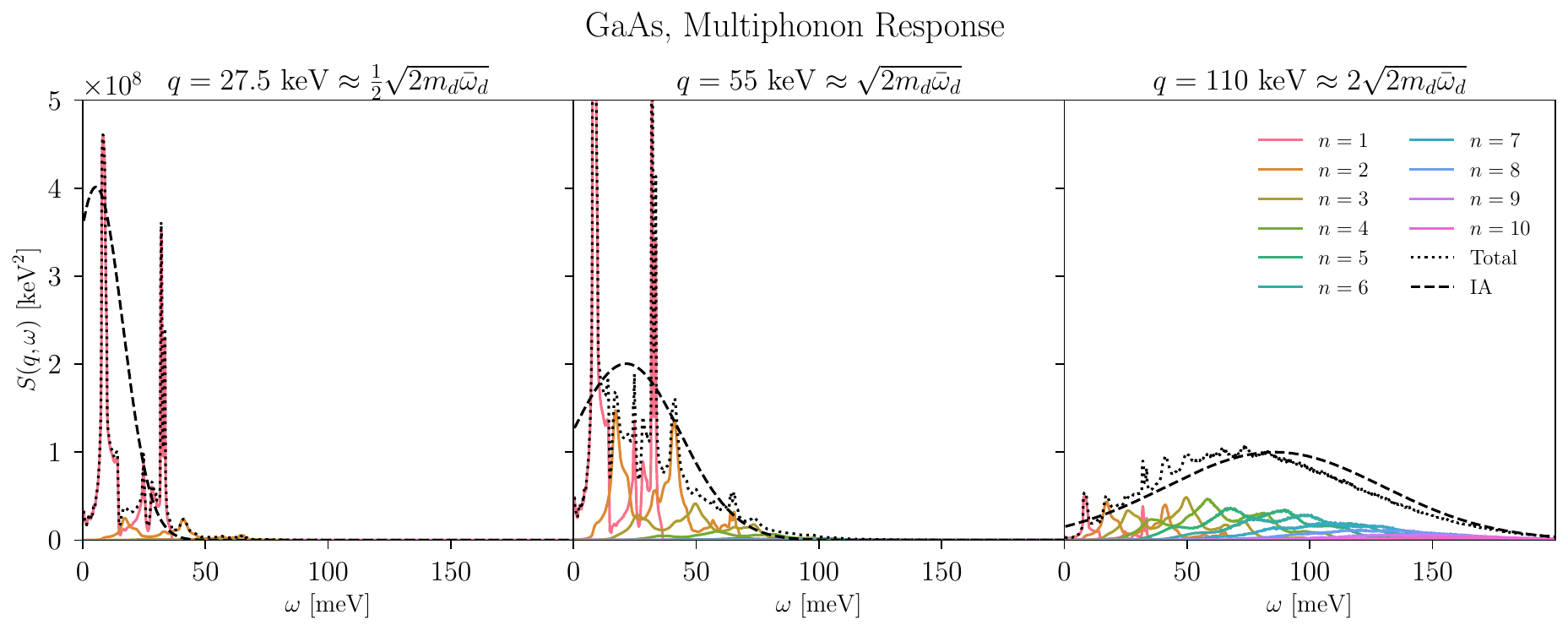}\hspace{0.2cm}} 
\newline
\centering
\subfloat[Cross sections for 3 events/kg-yr in GaAs for a hadrophilic mediator. Rates are computed with the $n\leq 10$ phonon terms in the incoherent approximation (solid lines), the impulse approximation (IA; dashed), and the analytic free nuclear recoil result (NR; dotted). We see that at sufficiently high masses--and hence momentum transfers--the impulse approximation sufficiently recovers the result of summing the phonon terms. Likewise, for yet larger momenta the impulse approximation merges onto the free nuclear recoil result, as discussed in Sec. \ref{sec:impulse_approx}.]{\label{fig:impulse-structures_bottom} 
\includegraphics[width=0.48\linewidth]{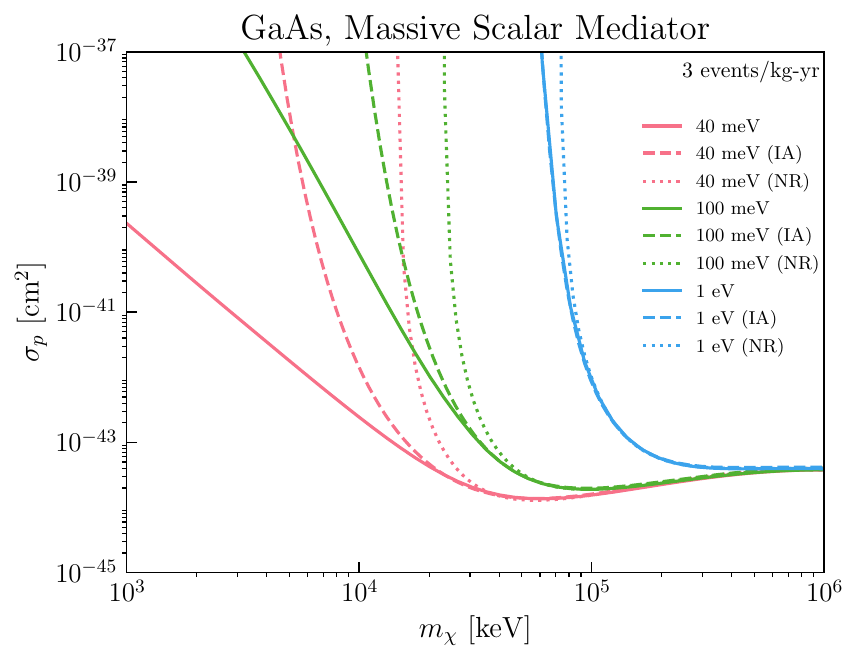} 
\includegraphics[width=0.48\linewidth]{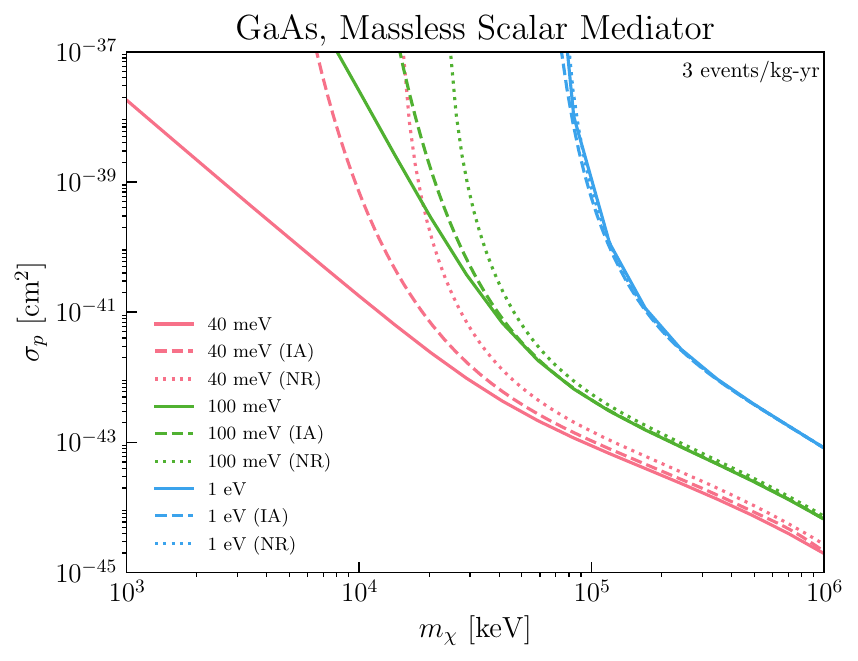} }
\caption{\textbf{Multiphonon transition into the nuclear recoil regime.}}
\label{fig:impulse-structures} 
\end{figure*}

In the previous section, where we dealt with $q<q_{\rm BZ}$, the incoherent approximation should be viewed as an order-of-magnitude estimate only. For $q>q_{\rm BZ}$, it is however on firm ground \cite{Schober2014,PhysRev.82.392} and is used routinely to measure the density of states from neutron scattering data \cite{Schober2014}. Moreover, in the $q\gg q_{\rm BZ}$ regime multiphonon processes become important. This follows from the form of the structure factor, obtained by inserting \eqref{eqn:n-incoherent} into the incoherent approximation \eqref{eq:inc-approx}:
 \begin{align}
    \label{eqn:n-incoherentstructure}
       S(\bfq, \omega) &\approx \frac{2 \pi}{\Omega_c} \sum_{d}^{\mathfrak{n}} (\overline{f_{d}})^2   e^{-2 W_d(\mathbf{q})}\sum_n   \left(\frac{q^2}{2 m_d} \right)^n  \nonumber \\ \times & \frac{1}{n!} \left(\prod\displaylimits_{i=1}^n \int d \omega_{i} \frac{D_d(\omega_i)}{\omega_i} \right) \delta \left(\sum_j \omega_j - \omega \right).
 \end{align}
From the discussion around \eqref{eq:typical-n}, the typical number of phonons is $n \sim \frac{q^2}{2 m_d \bar \omega_d}$. With \mbox{$\bar \omega_d\gtrsim 30$ meV} and \mbox{$m_d\gtrsim 30$ GeV} for most crystals, the self-consistency condition for the incoherent approximation ($q\gtrsim q_{\rm BZ}$) is therefore always satisfied for $n>2$ processes. The evolution of \eqref{eqn:n-incoherentstructure} for increasingly large $q$ is shown in Fig.~\ref{fig:impulse-structures_top}.

We can obtain an approximate scaling for \eqref{eqn:n-incoherentstructure} by separating each term in the sum over $n$ into $q$-dependent and $\omega$-dependent parts. The $\omega$-dependent part is given by the second line of the equation, which is only non-zero at $\omega \lesssim n \, \omega_{\mathrm{LO}}$ in order to satisfy the delta function. This part of the structure factor can be estimated to have at most the value of $1/(n! \, \bar \omega_d^{n+1})$; this is illustrated in Fig.~\ref{fig:n-phonon-omega-dependence} of Appendix~\ref{app:results_darkelf}, where we plot the numerical result. 
For $q\lesssim \sqrt{2m_d\bar \omega_d}$ (left and center panels of Fig.~\ref{fig:impulse-structures_top}), the Debye-Waller factor can be neglected and the structure factor then scales as $S(q,\omega) \propto \sum_n \frac{1}{n!} \left(\frac{q^2}{2 m_d \bar \omega_d}\right)^n$.  For $q^2/(2 m_d \bar \omega_d)\lesssim 1$, the structure factor therefore scales as $S(q,\omega) \sim q^{2m}$, with $m$ the lowest number of phonons that is kinematically allowed. This scaling will be useful in Sec.~\ref{sec:results}, where we use it to extract the approximate scaling behavior of the DM cross section curves. It no longer holds for $q\gtrsim \sqrt{2m_d\bar \omega_d}$ (right-hand panel of Fig.~\ref{fig:impulse-structures_top}), where many modes contribute equally. This regime however can be understood in the impulse approximation, which is the subject of the next section.


\subsection{The impulse approximation ($q\gg q_{\rm BZ}$)}
\label{sec:impulse_approx}

For $q\gg q_{\rm BZ}$ the sum of the multiphonon terms asymptotes to an approximately Gaussian envelope, as can be seen most clearly from the rightmost panel in Fig.~\ref{fig:impulse-structures_top}. This asymptotic form can be derived directly with a steepest descent approximation, also known as the \emph{impulse approximation}. It is valid whenever the interaction with the probe particle happens on a time scale short compared to that of the phonon modes.

To derive this, it is most insightful to take a step back from \eqref{eqn:n-incoherentstructure} and return to using \eqref{eq:correldensitystates} in \eqref{eqn:incohmigdal}. The auto-correlation function is then
\begin{equation}
    \label{eqn:incoh}
    \mathcal{C}_{\bfl d} = \frac{1}{V} e^{-2 W_d(\mathbf{q})}  \int\displaylimits_{-\infty}^{\infty} \! dt\, e^{ \frac{q^2}{2  m_d} \int d \omega^\prime \, \frac{D_d(\omega^\prime)}{\omega^\prime} e^{i \omega^\prime t}} e^{-i\omega t}.
\end{equation}
When $q \gg \sqrt{2 m_d \bar \omega_d}$, the exponent involving the density of states integral will be highly oscillatory in $t$, and the integral may be approximated by expanding about $t=0$ through a steepest descent method. (See Appendix \ref{app:impulse}). Doing so gives 
\begin{equation}
\label{eq:impulse_approx_C}
    \mathcal{C}_{\bfl d} \approx \frac{1}{V} \sqrt{\frac{2 \pi}{\Delta_d^2}} \exp \left({- \frac{ \big(\omega - \frac{q^2}{2m_d} \big)^2}{2 \Delta_d^2}} \right)
\end{equation}
where $\Delta_d^2 \equiv \frac{q^2 \bar{\omega}_d}{2 m_d}$. This approximation is referred to as the impulse approximation since the saddle-point around $t=0$ dominates the rate. The true peak is shifted slightly from the result \eqref{eq:impulse_approx_C}, which can be corrected by including higher orders in the expansion~\cite{Liang:2022xbu}. Including these additional terms has negligible impact on later results.

From \eqref{eq:impulse_approx_C}, we see that the structure factor in the impulse approximation is
\begin{equation}
    S^{\mathrm{IA}}(q, \omega) = \sum\displaylimits_d^{\mathfrak{n}} \frac{(\overline{f_{d}})^2 }{\Omega_c} \sqrt{\frac{2 \pi}{\Delta_d^2}} \exp \left({- \frac{ \big(\omega - \frac{q^2}{2m_d} \big)^2}{2 \Delta_d^2}} \right)
    \label{eq:impulseapprox}
\end{equation}
which is a sum of Gaussians peaked around $q = \sqrt{2 m_d \omega}$, one for each atom in the unit cell. In Fig.~\ref{fig:impulse-structures_top} we see that \eqref{eq:impulseapprox} is a reasonable approximation for $q\approx \sqrt{2m_d \bar \omega_d}$ and converges rapidly to the full result in \eqref{eqn:n-incoherentstructure} for $q\gtrsim 2 \sqrt{2m_d \bar \omega_d}$. As expected, it does not capture the features in the structure factor for $q \lesssim \sqrt{2 m_d \bar \omega_d}$.  In our final results, we use \eqref{eq:impulseapprox} for $q > 2\sqrt{2 m_d \bar{\omega}_d}$, as it is numerically much faster than \eqref{eqn:n-incoherentstructure}. For crystals composed of multiple atoms, we define the boundary as $\text{max}_d \big[2\sqrt{2 m_d \bar{\omega}_d}\big]$. At this scale, the average number of phonons is about four, and it is sufficient to truncate the sum at $n=10$ for all smaller $q$.

As we consider larger DM masses which access larger $q$ and $\omega$, the Gaussian becomes more sharply peaked. This can be seen by comparing the width $\Delta_d$ to the peak value $\omega=q^2 / 2 m_d$. In the large-$q$ limit, we have
\begin{equation}
    \label{eq:delta-over-omega}
    \lim\displaylimits_{q \rightarrow \infty} \frac{\Delta_d}{\omega} \approx   \sqrt{\frac{ \bar{\omega}_d}{\omega}} 
\end{equation}
so the Gaussian becomes narrow for $\omega$ well above the typical phonon energy. Then the narrow width limit exactly reproduces the expected free nuclear recoil delta function response:
\begin{align}
    \lim\displaylimits_{q, \, \omega \rightarrow \infty} \mathcal{C}_{\bfl d} &= \frac{2 \pi}{V} \delta  \left( \omega - \frac{q^2}{2 m_d} \right) \\
    \label{eq:structure_factor_nuclear_recoil}
    S^{\text{FR}} (q, \omega) &= \sum\displaylimits_d \frac{2 \pi}{\Omega_c} (\overline{f_{d}})^2  \,\delta  \left( \omega - \frac{q^2}{2 m_d} \right).
\end{align}
We therefore recover the familiar free nuclear recoil response for each individual atom in the unit cell.


In Fig.~\ref{fig:impulse-structures_bottom} we show cross section curves with a GaAs target, for both massive and massless scalar mediators. We compare the reach obtained with the full structure factor (in the incoherent approximation), the impulse approximation, and the free nuclear recoil limit. For $m_\chi \lesssim 20-40$ MeV, the full structure factor must be used to capture the rate, depending on the mediator mass and threshold. For $m_\chi \gtrsim 20-40$ MeV, the $q$ values compatible with the impulse approximation start to dominate, and we see that it reproduces the full result very closely. At even higher masses, the free nuclear recoil response becomes an excellent approximation, as expected.

A particular feature to notice from Fig.~\ref{fig:impulse-structures_bottom} is that the free nuclear recoil rate agrees with the impulse approximation result even in regions of the $q, \, \omega$ phase space where the Gaussian is not narrow. For example, for the massive mediator and $m_\chi=50$ MeV, the rate will be dominated by momentum transfers $q\sim 2 m_\chi v \sim 100$ keV, corresponding most closely to the rightmost panel of Fig.~\ref{fig:impulse-structures_top}. From~\eqref{eq:delta-over-omega} this gives $\Delta_d/\omega \approx 0.5$ which is not particularly small. The nuclear recoil approximation nevertheless works remarkably well. The reason is that phase space integral in \eqref{eq:cross-sec-eqn} has a trivial $\omega$ dependence aside from the $S(\bfq,\omega)$ factor, since the delta function in $\omega$ just determines the region of phase space that is integrated over. Therefore, as long as the energy threshold is small compared to the peak in  $\omega$, the phase space integral over \eqref{eq:impulseapprox} and \eqref{eq:structure_factor_nuclear_recoil} yields similar answers.

\subsection{Summary}

\begin{figure}
\centering

\includegraphics[width=0.47\textwidth, valign=m]{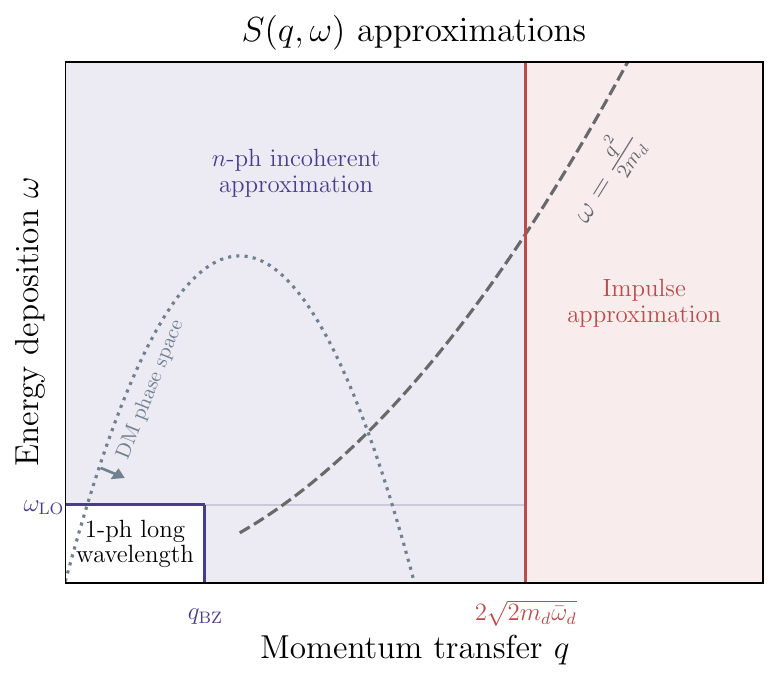}
\caption{Schematic figure (not to scale) depicting the approximation used to calculate the structure factor in various regions of phase space. The ``1-ph long wavelength'' regime is discussed in Sec.~\ref{sec:singlephonon}, the ``$n$-ph incoherent approximation" regime in Sec.~\ref{sec:twophonon} and \ref{sec:multiphonon} and the ``Impulse approximation'' region in Sec.~\ref{sec:impulse_approx}. 
 \label{fig:illustrative_plot}
}
\end{figure}

\label{sec:structuresummary}

Fig.~\ref{fig:illustrative_plot} schematically illustrates the various approximations for the structure factor discussed in this section. The boundaries reflect only our choice of approximation and not a sharp transition in the behavior of the structure factor. The dotted gray parabola represents the phase space boundary for a given $m_\chi$ and $v$ (see Sec.~\ref{sec:results}). This parabola extends upwards and rightwards as $m_\chi$ is increased, such that multiple different regimes are sampled for high enough $m_\chi$.

\begin{figure}
\centering

\includegraphics[width=\linewidth]{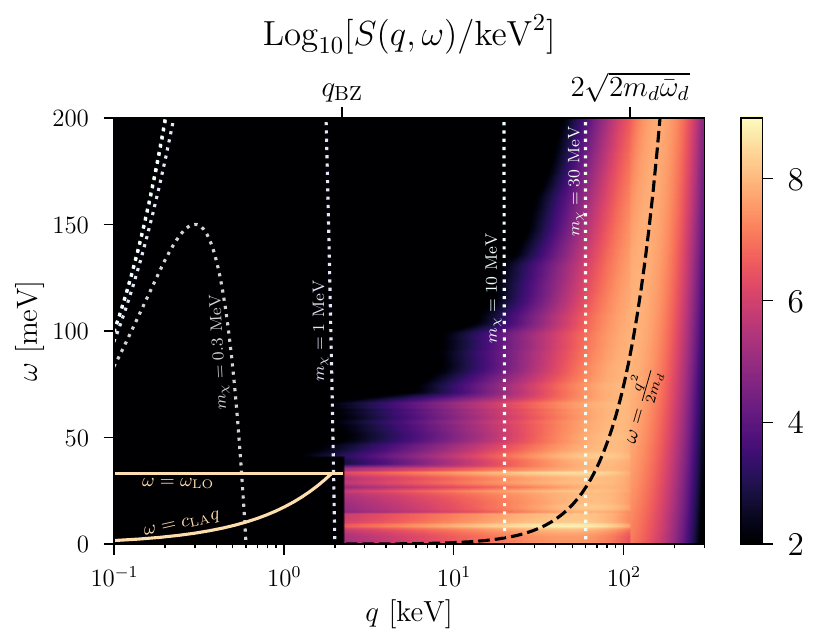}
\caption{{\bf GaAs structure factor}. Density plot of the structure factor in the same regimes of ($q,\omega)$ as shown in Fig.~\ref{fig:illustrative_plot}. Dotted lines are phase space boundaries for various DM masses with a typical initial velocity $v=10^{-3}.$ At low $q$ and $\omega$, the solid yellow lines are the dispersion relations of the single LA and LO phonons. At large $q$, the black dashed line is the free nuclear recoil dispersion relation; in general, there are separate lines for Ga and As but for clarity we show only one line corresponding to the average mass of Ga and As.  
\label{fig:GaAs_density_plot}}
\end{figure}

For the single phonon excitations $(n=1)$ described in Sec.~\ref{sec:singlephonon}, we use the long-wavelength and incoherent approximations for $q<q_{\rm BZ}$ and $q>q_{\rm BZ}$, respectively. This combination gives good agreement with a full DFT calculation of the scattering rate, at least for a cubic crystal such as GaAs.

For multiphonon excitations ($n\geq2$), we use the incoherent approximation for the structure factor for all $q$ below $\text{max}_d[2\sqrt{2 m_d \bar \omega_d}]$. This is motivated by Sec.~\ref{sec:twophonon}, where we argued that the incoherent approximation can serve as an order-of-magnitude estimate even for $q \ll q_{\rm BZ}$. Given the limitations of the long-wavelength approximation, a dedicated DFT calculation is needed in this regime. For multiphonon excitations, we sum terms in \eqref{eqn:n-incoherentstructure} until we achieve convergence, as explained in Sec.~\ref{sec:multiphonon}. Finally, for $q \ge \text{max}_d[2\sqrt{2 m_d \bar \omega_d}]$ we make use of the impulse approximation, which ultimately transitions into the well-known free nuclear recoil regime. This was explained in Sec.~\ref{sec:impulse_approx}.
 
Fig.~\ref{fig:GaAs_density_plot} shows our full calculation of the structure factor for GaAs, overlaid with the phase space boundaries for a few representative DM masses. In the low $q$, single phonon regime, the response is given by a set of $\delta$-functions on the LO and LA phonon dispersions, represented by the orange curves. At intermediate and high $q$, the structure function is modeled by a continuous function, where the layered structure for $\omega \lesssim 50$ meV reflects the various single and multiphonon contributions. At higher $q$ and $\omega$ the individual resonances cease to be visible and one transitions into the smooth $S(\bfq,\omega)$ predicted by the impulse approximation. At very high $\omega$ the structure function converges towards its free nuclear recoil form, which is represented by the black dashed line.

\begin{figure*}
\centering
\includegraphics[width=0.47\linewidth]{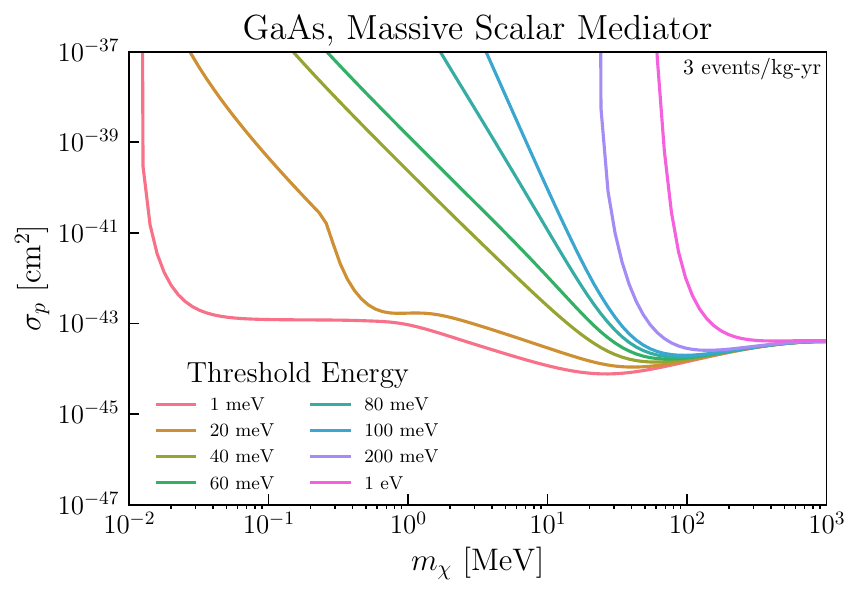} 
\includegraphics[width=0.47\linewidth]{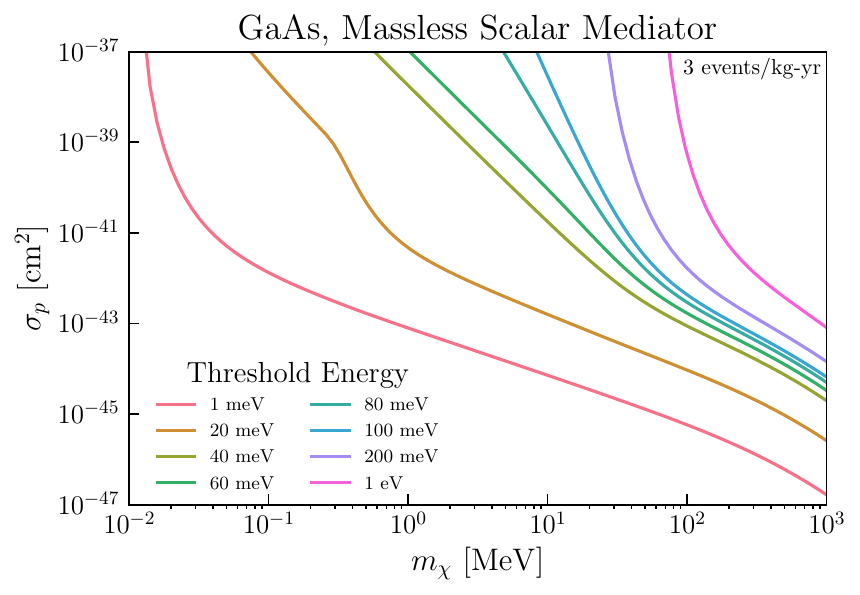} 
\caption{ Cross section plots corresponding to a rate of 3 events/kg-yr for massive and massless scalar mediators in GaAs for various thresholds. The structure factors used are the analytic results demarcated in Fig.~\ref{fig:illustrative_plot} for each corresponding regime in the $(q, \omega)$ phase space. For the massive mediator, we see the dominance of the single acoustic phonon at low masses and low thresholds, and of the optical phonon for intermediate thresholds. Eventually, for sufficiently high masses the process becomes dominated by the free nuclear recoil response. For the massless mediator, the $q^{-4}$ form factor favors small momenta, and the rate is dominated by the lowest accessible mode for a given threshold. 
\label{fig:GaAs_reaches}}
\end{figure*}

\section{Results} 
\label{sec:results}
In this section we convert our newly-gained understanding of the structure factor into concrete predictions for the DM scattering rate in a crystal target.
The event rate per unit of target mass is
\begin{align}
    \label{eq:rate-notisotropic}
    R  &= \frac{1}{\sum\displaylimits_d m_d} \frac{\rho_\chi}{m_\chi} \int d^3 \bfv\, v f(\bfv) \int d^3 \bfq\, d\omega \frac{d\sigma}{d\bfq d\omega}
\end{align}
where the experimental energy threshold is implicit in the boundary of the $\omega$ integral. $f(\bfv)$ is the DM velocity distribution, which we take to be
\begin{align}
&f\left(\bf{v}\right) = \frac{1}{N_0} \exp \left[-\frac{\left(\mathbf{v}+\mathbf{v}_{e}\right)^{2}}{v_{0}^{2}}\right] \Theta\left(v_{e s c}-\left|\mathbf{v}+\mathbf{v}_{e}\right|\right) \,,\nonumber \\
&N_0 = \pi^{3 / 2} v_{0}^{3}\left[\operatorname{erf}\left(\frac{v_{e s c}}{v_{0}}\right)-2 \frac{v_{e s c}}{v_{0}} \exp \left(-\frac{v^{2}_{e s c}}{v_{0}^{2}}\right)\right] \,,
\end{align}
with $v_0=220$ km/s, the Earth's average velocity $v_e = 240$ km/s, and $v_\text{esc}=500$ km/s the approximate local escape velocity of the Milky Way. The scattering rate can be further simplified in the isotropic limit; using \eqref{eq:cross-sec-eqn},
\begin{equation}
    \label{eq:rate-isotropic}
    R =  \frac{1}{4\pi \rho_T} \frac{\rho_\chi}{m_\chi}  \frac{\sigma_p}{\mu_\chi^2}  \int \! d^3 \bfv\, \frac{f(\bfv)}{v} \int\displaylimits_{q_-}^{q_+} \! dq \,  \int\displaylimits_{\omega_\mathrm{th}}^{\omega_+} \! d\omega \,   q \, |\Tilde{F}(q)|^2 S(q, \omega)
\end{equation}
where $\omega_\mathrm{th}$ is  the energy threshold of the experiment, and the other integration limits\footnote{In numerical implementations of \eqref{eq:rate-isotropic}, as done in \texttt{DarkELF}, it is beneficial to change the order of integration by first integrating over $v$, then $q$, and finally over $\omega$.} are 
\begin{align}
q_\pm &\equiv m_\chi v \left(1\pm\sqrt{1-\frac{2\omega_\mathrm{th}}{m_\chi v^2}}\right)\label{eq:qphasespace}\\
\omega_+ &\equiv qv -\frac{q^2}{2 m_\chi}. \label{eq:omegaphasespace}
\end{align} 
Note \eqref{eq:omegaphasespace} defines the phase space boundary shown in Fig.~\ref{fig:illustrative_plot} for a given $m_\chi$ and $v$.
Finally, $\rho_T$ is the mass density of the target material and we have recast the rate in terms of the DM-proton scattering cross section $\sigma_p \equiv 4 \pi b_p^2$.

\subsection{Massive hadrophilic mediator}
\label{subsec:massive_mediator}

In the case of a massive mediator coupling to baryon number, we calculate the scattering rate by taking \mbox{$\overline{f_d} = A_d$} and \mbox{$\tilde F(q) = 1$}. The cross sections corresponding to a rate of 3 events/kg-year exposure are shown in the left panel of Fig.~\ref{fig:GaAs_reaches}, assuming a GaAs target and for different energy thresholds. The same figures for Si, Ge and diamond can be found in Appendix~\ref{app:extraresults}.

We can understand the numerical results in Fig.~\ref{fig:GaAs_reaches} analytically using the scaling of the structure factor discussed in Secs.~\ref{sec:singlephonon}--\ref{sec:impulse_approx}. 
First, from \eqref{eq:rate-isotropic}, the $m_\chi$ dependence of the rate is contained in 
\begin{equation}
    \label{eq:massive_mediator_rate}
    R \propto \frac{\sigma_p}{m_\chi \mu_\chi^2}  \int\displaylimits_{q-}^{q+} \! d q \int\displaylimits_{\omega_\mathrm{th}}^{\omega_+} \! d\omega \, q \, S(q, \omega).
\end{equation}
The structure factor only contains positive powers of $q$ across the entire phase space, so for a massive mediator, the integral \eqref{eq:massive_mediator_rate} will be dominated by the largest kinematically accessible momentum transfers.

For $m_\chi \gg 30$ MeV, the kinematically allowed phase space is extended to $q$ and $\omega$ where the free nuclear recoil approximation can be used. The rate therefore approximately scales as $R\sim 1/m_\chi$ for $m_p\gtrsim m_\chi \gg$ 30 MeV. For low enough thresholds, this scaling holds  even as the dark matter mass comes within $O(\mathrm{few})$ of $30$ MeV, where the structure factor is relatively broad in $\omega$. 
The reason is that the kinematically allowed phase space is wide enough in $\omega$ that the integral over the Gaussian in the impulse approximation gives within a factor of few of the integral over the delta function in \eqref{eq:structure_factor_nuclear_recoil}, as discussed earlier in Sec.~\ref{sec:impulse_approx}.

For dark matter masses of 1 to 30 MeV, the allowed phase space is restricted to values of $q < \sqrt{2 m_d \bar{\omega}}$. Here the structure factor can be expanded in powers of $q/\sqrt{2 m_d \bar{\omega}}$ and favors small $\omega$. As noted in Sec.~\ref{sec:multiphonon} the structure factor scales as $\sim q^{2m}$, with $m$ the smallest number of phonons whose total energy is above the energy threshold. We see there is significant threshold dependence: the single phonon final state strongly dominates the rate if it is above the energy threshold, while for higher thresholds only multiphonons contribute. The rate integral now scales as
\begin{equation}
    \label{eq:rate_scaling_massive_nphonon}
    R \propto \frac{\sigma_p}{m_\chi^3} \! \! \int\displaylimits^{2 m_\chi v} \!\!\! dq \, q^{2m+1} \! \! \! \int\displaylimits_{\omega_{\mathrm{th}}} \! d \omega  \propto \sigma_p \, m_\chi^{2 m -1},
\end{equation}
where $q$ was evaluated at its maximum $q \sim 2 m_\chi v$. The $\omega$ integral does not contribute to the $m_\chi$ scaling of the rate, since the integrand is peaked in $\omega$ somewhere near the energy threshold $\omega_\mathrm{th}$. This expression then gives the approximate scaling $R \propto m_\chi^{2m - 1}$. Since $m$ is dependent on the energy threshold, this explains why different thresholds in Fig.~\ref{fig:GaAs_reaches} result in a different scaling as a function of $m_\chi$.  

At even lower dark matter masses ($m_\chi < 1$ MeV), the phase space is restricted to $q$ values within the first Brillouin zone, which is dominated by single phonon production in the long wavelength regime. If the threshold is low enough to access a single phonon, the scaling further depends on whether the threshold captures an appreciable part of the LA branch. If so, the leading contribution comes from the acoustic mode \eqref{eq:single-ph-aco-analytic}, which gives
\begin{equation}
    R \propto \frac{\sigma_p}{m_\chi^3}\! \!\int\displaylimits^{2 m_\chi v} \! \! \! dq \, q^2 \! \int d\omega \,\delta(\omega - c_{\mathrm{LA}} q) \propto \sigma_p,
\end{equation}
approximately independent of $m_\chi$. This behavior is clearly reproduced in Fig.~\ref{fig:GaAs_reaches} for the 1 meV threshold, for which the acoustic branch is always accessible. If the threshold is too high to access the acoustic branch, but can detect the optical branch, the structure factor has an extra $q^3$ scaling and we find  $R \propto m_{\chi}^{3}$.  This case occurs for $m_\chi\lesssim$ 0.3 MeV on the 20 meV curve in Fig.~\ref{fig:GaAs_reaches}. For $m_\chi\gtrsim$ 0.3 MeV the DM can excite the acoustic branch, resulting in a sharp enhancement of the rate.

\subsection{Massless hadrophilic mediator}
\label{subsec:IVB_massless_mediator}

If we instead have a massless mediator that couples to baryon number, then by convention, the mediator form factor is taken to be $|\tilde F(q)|^2 =  \big( \frac{m_{\chi} v_0}{q} \big)^4$ with $v_0 = 220$ km/s. The cross section  curves for this scenario are given in the right panel of Fig.~\ref{fig:GaAs_reaches} again for different thresholds.

As in Sec.~\ref{subsec:massive_mediator}, we can analytically explain the scaling of the different curves across the DM mass range. The main difference with the massive mediator case is that for a massless mediator, there is a $1/q^4$ scaling in the form factor, which leads to a scattering rate that generally favors low $q$ and $\omega$. The main contribution to the rate will therefore be much more threshold dependent across all DM masses.

If the threshold is small enough to access single acoustic phonon excitations, then this will be the dominant contribution to the rate at all masses. Again from \eqref{eq:rate-isotropic} and using the analytic acoustic structure factor, the rate for thresholds that are sensitive to a single acoustic phonon scales as
\begin{equation}
    \label{eq:rate-acoustic}
    R \propto \sigma_p \, m_\chi \!\!\! \!\!\! \int\displaylimits_{\omega_\mathrm{th}/c_{\text{LA}}} \! \!\!\! \!\!\! dq \, \frac{1}{q^2} \! \int \!  d\omega  \, \delta (\omega - c_\mathrm{LA} q).
\end{equation}
The integrand is largest at the smallest $q$, so we estimate the $q$ integral by evaluating the integrand at $q\approx \omega_\mathrm{th}/c_{\text{LA}}$ in \eqref{eq:qphasespace}. The integrand therefore has no $m_\chi$ dependence and gives the  scaling $R\propto m_\chi$  for the $\omega > 1$ meV curve in Fig.~\ref{fig:GaAs_reaches}. Note however that this scaling behavior is sensitive to our convention for the reference momentum in $\tilde F(q)$. For example, in models with both electron and nucleon couplings one often chooses to normalize the form factor with the reference momentum $q_0=\alpha m_e $, which would yield  $R\propto m_\chi^{-3}$.

If the LA branch is not accessible but the LO branch is, the production of a single LO mode will generally dominate. This introduces a different $m_\chi$ dependence, which can be seen in Fig.~\ref{fig:GaAs_reaches} by comparing the 1 meV and 20 meV curves in the region with $m_\chi\lesssim 30$ MeV. If $m_\chi < 1$ MeV, using the expression in \eqref{eq:single-ph-opt-analytic} gives
\begin{equation}
    \label{eq:rate_optical_sub_MeV}
    R \propto \sigma_p m_\chi  \! \! \! \int\displaylimits^{2 m_\chi v} \!\!\! dq \, q \int \! d\omega \,  \delta (\omega - \omega_{LO}).
\end{equation}
Unlike for the acoustic phonon, the structure factor favors high $q$ so that the largest contribution is near $q \sim 2 m_\chi v$, giving $R \propto m_\chi^{3}$. If $m_\chi > 1$ MeV, the rate integrand is dominated by momentum transfers $q \sim q_\mathrm{BZ}$. This is because when $q > q_\mathrm{BZ}$ and $\omega \le \omega_\mathrm{LO}$ we are using the incoherent approximation for single phonon production, where the $q$ integrand drops as $q^{-1}$. Thus, we estimate the rate by integrating up to $q_{\rm BZ}$ only:
\begin{equation}
    R \propto \sigma_p m_\chi  \! \int\displaylimits^{q_{\mathrm{BZ}}} \! dq \, q \! \int \! d \omega \,  \delta (\omega - \omega_{LO}),
\end{equation}
and find that $R \propto m_\chi$. This is the reason why the 20 meV curve in Fig.~\ref{fig:GaAs_reaches} changes slope around $m_\chi\sim 1$ MeV.

We next turn to the intermediate mass range ($1-30$ MeV) with $\omega_\mathrm{th} > \omega_\mathrm{LO}$, such that $n \ge 2$ phonons. In Fig.~\ref{fig:GaAs_reaches} this corresponds to the curves with thresholds of 40 meV and above. As in Sec.~\ref{subsec:massive_mediator}, we again notice that the leading contribution to the structure factor will be given by the smallest number of phonons, $m$, that can exceed the threshold energy. In this regime, the integrand $\propto S( q, \omega)/q^3$ scales with positive powers of $q$ for $m\ge2 $ phonons, since \eqref{eqn:n-incoherent} grows faster than $q^3$. The analysis for multiphonons then follows exactly the same logic as the discussion in the previous section and we find that $R \propto m_\chi^{2m - 1}$.

For large dark matter masses ($\gg 30$ MeV), again if the threshold is well above the single phonon energy, we can apply the free nuclear recoil approximation to obtain the scaling. Using the free nuclear structure factor gives
\begin{equation}
R \propto \frac{\sigma_p}{m_\chi^3} \!\!  \int\displaylimits_{\sqrt{2 m_d \omega_\mathrm{th}}}\!\!\!\!\!\!\!\! dq \, q \Big( \frac{m_\chi v_0}{q} \Big)^4  \int\! d \omega\, \delta \big (\omega - \frac{q^2}{2 m_d} \big).
\end{equation}
The $q$-integral is dominated by low-momentum transfers along the free nuclear recoil dispersion, so we evaluate the integral at the intersection of $\omega = \omega_\mathrm{th}$ and $\omega = \frac{q^2}{2m_d}$, or $q = \sqrt{2m_d \omega_\mathrm{th}}$. Then, the approximate scaling in this regime is  $R \propto m_{\chi}/\omega_{\mathrm{th}}$, which we verify numerically in Fig.~\ref{fig:GaAs_reaches}.

\subsection{Dark photon mediators}

\begin{figure}
\includegraphics[width=0.45\textwidth]{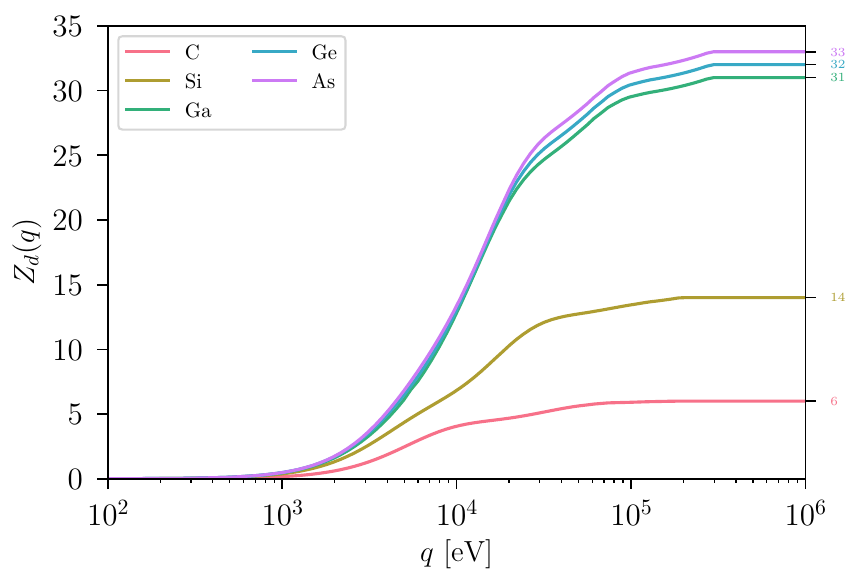}
\caption{Momentum dependence of the effective ion charge for atomic elements, as computed in \cite{BrownXray}. \label{fig:zeff}}
\end{figure}

The defining feature of a dark photon mediator is that it couples to the electric charge of the SM particles.  In the regime where phonons are the relevant degrees of freedom, the charge of the nucleus is (partially) screened by the electrons. This means that we need a notion of an \emph{effective charge}, as seen by the DM, which is momentum dependent. For individual atoms, this effective charge interpolates between zero in the low momentum, fully screened regime and the nuclear charge in the high momentum regime. We use the calculations from Brown~et.~al. \cite{BrownXray} of the effective charge for individual atoms, as shown in Fig.~\ref{fig:zeff}.  We expect this approximation to hold only for $q \gtrsim q_{\rm BZ}$, since additional many-body effects should be relevant for $q < q_{\rm BZ}$. This is particularly true for a polar material such as GaAs, where the Born effective charge of the Ga and As atoms is non-zero in the $q\to 0$ limit. In this regime a full DFT calculation of the momentum dependence of the effective charge is needed, which we do not attempt here.

In this work, we will therefore focus on the  momentum regime $q \gtrsim q_{\rm BZ}$, which corresponds to \mbox{$m_\chi \gtrsim $ MeV}. In this case we can use the incoherent approximation and take \mbox{$\overline{ f_d}=Z_d(q)$}, with $Z_d(q)$ the atomic effective charges in Fig.~\ref{fig:zeff}. This allows us to compute scattering rates with dark photon mediators for the production of two or more phonons, which is dominated by the highest kinematically accessible momentum transfers.

\begin{figure*}
\centering
\includegraphics[width=0.47\linewidth]{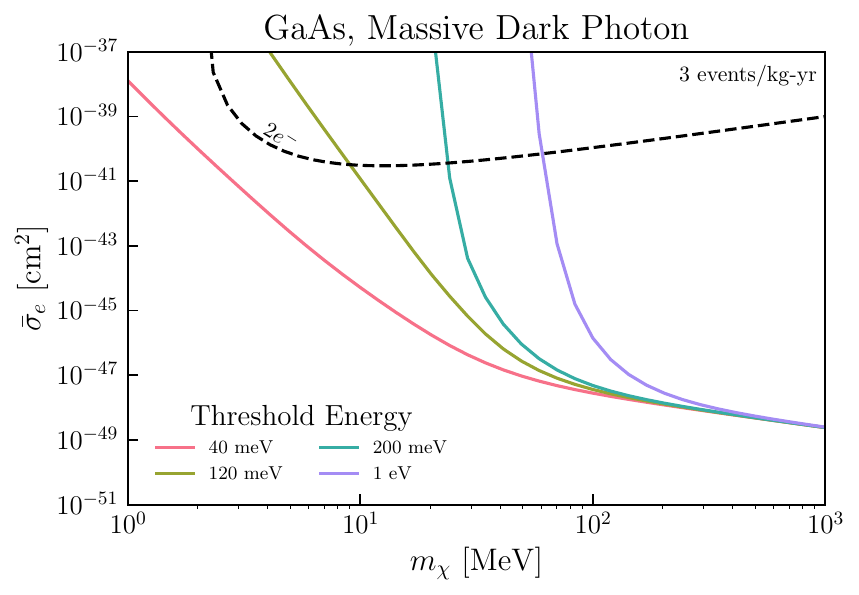} 
\includegraphics[width=0.47\linewidth]{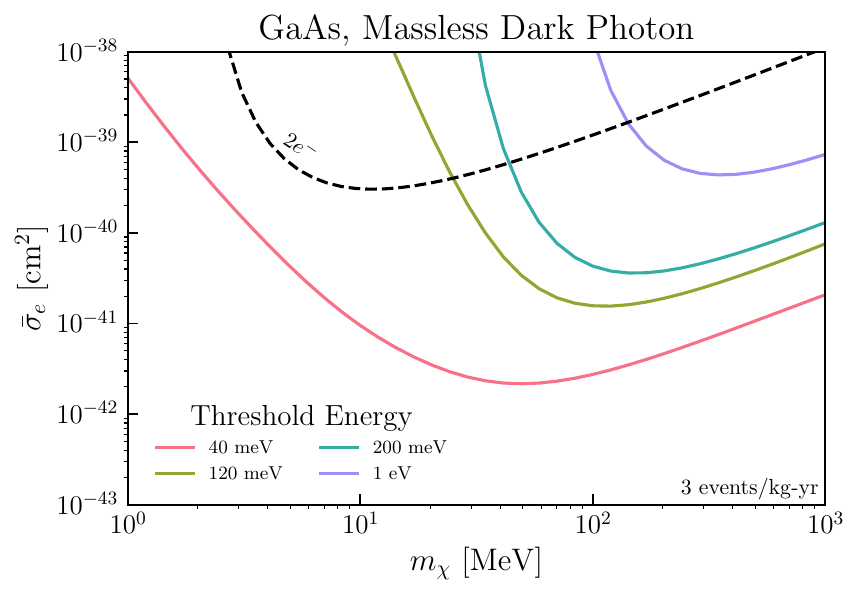}

\caption{ Cross section plots for a rate of 3 events/kg-year in GaAs, for massive and massless dark photon mediators. For comparison, the dashed black lines represent the cross sections required for DM-electron scattering with a 2$e^-$ ionization threshold with the same exposure, as computed using \texttt{DarkELF}~\cite{Knapen:2021run,Knapen:2021bwg}.  
\label{fig:reaches_dark_photon}}
\end{figure*}

The regime $q < q_{\rm BZ}$ is relevant primarily for massless dark photon mediators. (For massive dark photon mediators, there are strong BBN  constraints that severely limit the scattering rate for sub-MeV dark matter, see e.g.~\cite{Knapen:2017xzo}.)
In this regime, there are substantial deviations from the atomic effective charges due to the delocalized nature of the valence electrons. For instance, a polar material such as GaAs, SiC and sapphire can have a residual dipole moment associated with atomic displacements even for $q\to 0$. The effective couplings $\bar f_d$ in this limit are given  by $Z_d^*/\epsilon_\infty$, where $Z_d^*$ is the Born effective charge and $\epsilon_\infty$ is a screening due to valence electrons; the Born effective charges can be calculated with DFT methods~\cite{Griffin:2018bjn,Griffin:2019mvc,Griffin:2020lgd}. This was treated in previous studies of single-phonon production through a massless dark photon mediator~\cite{Knapen:2017ekk,Griffin:2018bjn,Trickle:2019nya,Cox:2019cod,Griffin:2019mvc,Trickle:2020oki,Griffin:2020lgd,Coskuner:2021qxo}.
For non-polar materials such as Si, Ge and diamond, the Born effective charges vanish and instead multiphonon production is expected to dominate. This can be estimated with the energy loss function~\cite{Knapen:2021bwg}, at least for sub-MeV dark matter. Since this $q < q_{\rm BZ}$ regime is already included in \texttt{DarkELF}~\cite{Knapen:2021bwg}, we restrict our results here to  multiphonon processes with $q>q_\mathrm{BZ}$ and $\omega > \omega_{\mathrm{LO}}$.

Our results are shown in Fig.~\ref{fig:reaches_dark_photon} for GaAs; the results for Ge, Si and diamond are deferred to Appendix~\ref{app:extraresults}. As is conventional for dark photon mediators, we choose the reference momentum for the massless mediator to be $q_0=\alpha m_e$ and present the results in terms of the effective DM-electron cross section $\bar \sigma_e$ \cite{Essig:2015cda}, with \begin{equation}
    \bar \sigma_e = \frac{\mu_{\chi e}^2}{\mu_\chi^2} \sigma_p
\end{equation}
and $\mu_{\chi e}$ the DM-electron reduced mass. In our calculations using the atomic effective charges, we impose $q>q_{\rm BZ}$ to ensure we are not sampling the area of phase space for which these charges are clearly invalid. This means that our rate calculations for $m_\chi\lesssim 10$ MeV are a slight underestimate of the true result. 
\section{Conclusions and outlook}
\label{sec:discuss}

It is well-known that DM scattering in crystals can lead to one or more phonons being produced if DM has MeV-scale mass, as well as a recoiling nucleus if DM has GeV or higher mass. These processes are two sides of the same coin, depending on whether the momentum transfer is comparable to the inverse of the interparticle spacing and whether the energy deposition is comparable to the typical phonon energy $\sim \bar \omega$. When both momentum and energy scales are small, single phonon production dominates, and when both are large, nuclear recoils dominate. Here we studied the intermediate regime which is dominated by many phonons, which allows us to smoothly interpolate between single phonon production and nuclear recoils (see Fig.~\ref{fig:GaAs_reaches}).

To make the multiphonon calculation tractable, we relied on the isotropic, incoherent, and harmonic crystal approximations. This allowed us to obtain analytic results for the scattering rate in terms of the phonon density of states in the crystal. These approximations are expected to be very good for $q\gg q_{\rm BZ}$ ($m_\chi\gg 1$ MeV), as they explicitly reproduce the nuclear recoil limit when $q \gg \sqrt{2 m_N \bar \omega}$. For $q\lesssim q_{\rm BZ}$ ($m_\chi\lesssim 1$ MeV) the experimental threshold determines which theoretical treatment is most appropriate: for single phonon production, one can obtain analytic formulas by instead using a long wavelength, isotropic approximation. These results are currently only valid for cubic crystals such GaAs, Si, Ge and diamond. For strongly anisotropic materials such as sapphire, one must find a way to generalize them further or rely on DFT calculations. For multiphonon production and $q \lesssim q_{\rm BZ}$, the situation is more complicated: in this case it cannot be taken for granted that anharmonic corrections to the various multiphonon channels can be neglected. The anharmonic multiphonon contributions involving optical modes are particularly difficult to model analytically, and at the moment we perform a simple estimate in a toy model to justify extrapolating the incoherent and harmonic approximations to $q \lesssim q_{\rm BZ}$. A dedicated DFT calculation is needed to improve their accuracy. 

Our approach provides a smooth description of sub-GeV dark matter scattering down to keV masses for hadrophilic mediators. For dark photon mediators, a DFT calculation of the momentum-dependent couplings in the $q \sim q_{\rm BZ}$ regime is needed to complete the interpolation. For both mediators, we have provided results for multiple direct detection materials of interest, and also included our calculation as part of the \texttt{DarkELF} public code package. These will be essential to interpret  direct detection results as experimental thresholds for calorimetric detectors reach the eV scale and lower.

\acknowledgments
We are grateful to So Chigusa, Sinéad Griffin, Bashi Mandava and Mukul Sholapurkar for useful discussions. 
BCD, TL, and EV were supported with Department of Energy grants DE-SC0019195 and DE-SC0022104, as well as a UC Hellman fellowship.  SK was supported by the Office of High Energy Physics of the U.S. Department of Energy under contract DE-AC02-05CH11231.
EV is supported by a Sloan Scholar Fellowship.
\appendix

\section{Two phonon analytic structure factors}
\label{sec:two_phonon_appendix}
In Sec.~\ref{sec:twophonon} we compared the long-wavelength and incoherent approximations for the two-phonon final states, for $q$ within the first BZ. In this appendix we provide the analytic expressions for both approximations.

\subsection{Long-wavelength approximation}

Here we discuss how we extend the analytic calculations from \cite{Campbell-Deem:2019hdx} for the coherent two-phonon structure factor to additional combinations of final state phonon pairs. As in Sec.~\ref{sec:twophonon}, we assume a hadrophilic mediator with $\overline{ f_d} = A_d$ throughout this appendix. It was shown in \cite{Campbell-Deem:2019hdx} that the structure factor separates into harmonic and anharmonic contributions 
\begin{equation}
    S(\bfq, \omega) = S^{\mathrm{(harm)}} (\bfq, \omega) + S^{\mathrm{(anh)}} (\bfq, \omega)
\end{equation}
which do not interfere at leading order in the long wavelength limit. The first term involves expanding \eqref{eq:coherent_previous} to second order; note that it was referred to as the contact term in \cite{Campbell-Deem:2019hdx}. The anharmonic term is computed using an anharmonic phonon interaction Hamiltonian to first order. The specific matrix elements to be used are given in equations (12) and (13) of~\cite{Campbell-Deem:2019hdx}. We take the long-wavelength approximation for the phonon modes, as described in Sec.~\ref{sec:singlephonon}. For a crystal with two atoms in the unit cell, the longitudinal eigenvectors can be approximated by 
\begin{align}
{\bf e}_\mathrm{LA,\bfk, 1} &\approx \frac{\sqrt{A_1}}{\sqrt{A_1 + A_2}} \hat{\bfk}, \\
{\bf e}_\mathrm{LA,\bfk,2} &\approx \frac{\sqrt{A_2}}{\sqrt{A_1 + A_2}} e^{-i \bfk \cdot \bfr_2^0} \hat{\bfk} \\
{\bf e}_\mathrm{LO,\bfk,1} &\approx \frac{\sqrt{A_2}}{\sqrt{A_1 + A_2}} \hat{\bfk}, \\ 
{\bf e}_\mathrm{LO,\bfk, 2} &\approx - \frac{\sqrt{A_1}}{\sqrt{A_1 + A_2}} e^{-i \bfk \cdot \bfr_2^0} \hat{\bfk}.
\end{align}
with $\hat{\bfk}$ the unit vector along the phonon propagation direction.
Note that the $\bfr_2^0$ dependence was neglected in the LA eigenvector in \eqref{eq:accoustic_eigenvectors} and in \cite{Campbell-Deem:2019hdx}; here we have kept this additional phase so that the acoustic and optical eigenvectors are explicitly orthogonal across a unit cell. This additional phase factor will only be relevant in cases where there is a destructive interference in the leading coupling to acoustic phonons, which occurs for some final states \cite{Cox:2019cod}. The transverse eigenvectors lay in the plane perpendicular to $\hat{\bfk}$ and have analogous normalizations. 

Analytic expressions for the harmonic structure factor were provided in Ref.~\cite{Campbell-Deem:2019hdx} for acoustic-acoustic final states only. We require expressions for the optical-optical and optical-acoustic final states as well to perform the comparison with the incoherent approximation. A straightforward application of (16) in \cite{Campbell-Deem:2019hdx} to the lowest order in $q$ gives
\begin{align}
\nonumber
S_\mathrm{LOLO}^\mathrm{(harm)} &=  \frac{2\pi}{\Omega_c} \frac{ \pi\, q^4 }{120  m_p^2 \omega_\mathrm{LO}^2 } \delta (\omega - 2 \omega_\mathrm{LO})
\\
\nonumber
S_\mathrm{LOTO}^\mathrm{(harm)} &= \frac{2\pi}{\Omega_c} \frac{\pi\, q^4}{90 m_p^2 \omega_\mathrm{LO} \omega_\mathrm{TO}  }  \delta (\omega - (\omega_\mathrm{LO} + \omega_\mathrm{TO} ))
\\
S_\mathrm{TOTO}^\mathrm{(harm)} &=  \frac{2\pi}{\Omega_c} \frac{\pi \, q^4}{45 m_p^2 \omega_\mathrm{TO}^2 } \delta (\omega - 2 \omega_\mathrm{TO})
\label{eq:coh-optical-optical}
\end{align}
for the optical-optical modes.

For the optical-acoustic modes, the harmonic structure factors are of the form 

\begin{align}
\label{eq:coh-optical-acoustic}
\nonumber
S_\mathrm{LOLA}^\mathrm{(harm)} = & \frac{2\pi}{\Omega_c} \frac{a^5 }{2304 \pi^2 c_\mathrm{LA}^2  m_p^2 \omega_\mathrm{LO} } \frac{A_1 A_2}{ (A_1 + A_2)^2 } \left(\frac{\omega - \omega_\mathrm{LO}}{c_\mathrm{LA}} \right)^7
\\
& \times 
g_\mathrm{LOLA}^\mathrm{(harm)} (x)\, \Theta ( c_\mathrm{LA} q_\mathrm{BZ} - (\omega - \omega_\mathrm{LO} ) ),
\end{align}
where 
$x \equiv \frac{c_\mathrm{LA} q}{\omega - \omega_\mathrm{LO}}$. The other structure factors for optical-acoustic final states are given by relabelings $\mathrm{LO} \rightarrow \mathrm{TO}$, $\mathrm{LA} \rightarrow \mathrm{TA}$, where the expressions $g$ expanded at small $q$ are 
\begin{align}
\nonumber
&g_\mathrm{LOLA}^\mathrm{(harm)} (x \ll 1) \approx \frac{3}{10} x^6 - \frac{1}{7} x^8 + \frac{1}{15} x^{10} \\
\nonumber
&g_\mathrm{LOTA}^\mathrm{(harm)} (x \ll 1) \approx \frac{1}{5} x^6 + \frac{12}{35} x^8 - \frac{4}{105} x^{10 }\\
\nonumber
&g_\mathrm{TOLA}^\mathrm{(harm)} (x \ll 1) \approx \frac{1}{5} x^6 + \frac{1}{7} x^8 - \frac{1}{15} x^{10} \\
&g_\mathrm{TOTA}^\mathrm{(harm)} (x \ll 1) \approx \frac{4}{5} x^6 - \frac{12}{35} x^8 + \frac{4}{105} x^{10}.
\end{align}
We see that at leading order in small $q$, the optical-acoustic structure factors are all suppressed by an additional factor of $q^2$ relative to the optical-optical modes, which is due to destructive interference. Since we will be comparing with the incoherent approximation at small $q$, we can effectively neglect these final states.

We would also like to compute the anharmonic contributions to the 2-phonon structure factor, which we do with the inclusion of an anharmonic interaction Hamiltonian. For acoustic phonons in the long-wavelength limit, we have an effective Hamiltonian for acoustic phonons where the  interactions are given in terms of  macroscopic properties of the crystal through the Lamé parameters, as described in~\cite{Campbell-Deem:2019hdx}. For the interactions of optical phonons, however, it is more difficult to write down a reliable analytic Hamiltonian. In this case we use (45) of Ref.~\cite{Campbell-Deem:2019hdx}, which comes from \cite{SRIVASTAVA1980357}. This Hamiltonian should be taken only at the order-of-magnitude level. We restrict the use of both effective Hamiltonians to the first BZ. The analytic expressions for the acoustic-acoustic and acoustic-optical final states are given already, so we complete this by calculating the optical-optical terms.  At leading order in $q$, this gives
\begin{align}
\label{eq:optical-optical-anharmonic}
\nonumber
S_\mathrm{LOLO}^\mathrm{(anh)} =& \frac{2\pi}{\Omega_c} \frac{\pi}{6 m_p^2} \frac{c_\mathrm{LA}^2}{\overline{c}^2} \frac{\omega_\mathrm{LO}^2 q^4} { ((2 \omega_\mathrm{LO})^2 - (c_\mathrm{LA} q)^2)^2 }
\\
&\times \delta(\omega-2\omega_{\text{LO}})\nonumber
\\
\nonumber
S_\mathrm{LOTO}^\mathrm{(anh)} =& \frac{2\pi}{\Omega_c} \frac{2 \pi}{3 m_p^2} \frac{c_\mathrm{LA}^2}{\overline{c}^2} \frac{\omega_\mathrm{LO} \omega_\mathrm{TO} q^4} { ((\omega_\mathrm{LO} + \omega_\mathrm{TO})^2 - (c_\mathrm{LA} q)^2)^2 }
\\
&\times \delta(\omega-\omega_{\text{LO}}-\omega_{\text{TO}})\nonumber
\\
S_\mathrm{TOTO}^\mathrm{(anh)} =& \frac{2\pi}{\Omega_c} \frac{2 \pi}{3 m_p^2} \frac{c_\mathrm{LA}^2}{\overline{c}^2} \frac{\omega_\mathrm{TO}^2 q^4} { ((2 \omega_\mathrm{TO})^2 - (c_\mathrm{LA} q)^2)^2 }\nonumber
\\
&\times \delta(\omega-2\omega_{\text{TO}}),
\end{align}
where $\overline{c}\equiv (c_\mathrm{LA}+c_\mathrm{TA})/2$. 
We have also assumed that the Gr\"{u}neisen constant $\gamma_\mathrm{G} \approx 1$.

\subsection{Incoherent approximation}

The second result needed for the comparison in Sec.~\ref{sec:twophonon} is the two-phonon structure factor for GaAs in the incoherent approximation. To calculate this, we use the simplified density of states in \eqref{eq:GaAs_toy_DoS} corresponding to the long-wavelength limit. Performing the $n=2$ integral in \eqref{eqn:n-incoherent} gives 
\begin{align}
    \label{eq:structure_n2_incoh}
    S_{n=2}(q,\omega) =  \mathcal{S}_\mathrm{LALA}+ \mathcal{S}_\mathrm{LATA} + \ldots 
\end{align}
where each $\mathcal{S}$ is a contribution to the $n=2$ structure factor from the part of the density of states associated with the subscripted modes, and the ellipsis indicates we sum over all combinations of modes. The first term of the sum in \eqref{eq:structure_n2_incoh} is
\begin{align}
    \label{eq:structure_n2_incoh_components}
    \nonumber
    \mathcal{S}_\mathrm{LALA} =& \frac{2\pi}{\Omega_c} \frac{q^4}{96 c_\mathrm{LA}^6 q_\mathrm{BZ}^6 m_p^2} \Bigg( \omega^3 \Theta( c_\mathrm{LA} q_\mathrm{BZ} - \omega ) \\
    \nonumber
    & - \left(4 c_\mathrm{LA}^3 q_\mathrm{BZ}^3 - 6 c_\mathrm{LA}^2 q_\mathrm{BZ}^2 \omega + \omega^3 \right) \\
    & \quad \times \Theta( \omega - c_\mathrm{LA} q_\mathrm{BZ}) \Theta (2 c_\mathrm{LA} q_\mathrm{BZ} - \omega) \Bigg),
\end{align}
and $\mathcal{S}_\mathrm{TATA}$ is given by $\mathcal{S}_\mathrm{LALA}$ with the replacement $\mathrm{LA} \rightarrow \mathrm{TA}$ and an additional overal factor of 4. The same procedure gives the $\mathrm{LATA}$ term as
\begin{align}
\nonumber
&\mathcal{S}_\mathrm{LATA} = \frac{2\pi}{\Omega_c} \frac{q^4}{24 c_\mathrm{LA}^3 q_\mathrm{BZ}^6 m_p^2}\  \Bigg( \frac{\omega^3}{ c_\mathrm{TA}^3} \Theta ( c_\mathrm{TA} q_\mathrm{BZ} - \omega) \\
\nonumber
& + \frac{-2 c_\mathrm{TA} q_\mathrm{BZ}^3 + 3 \omega q_\mathrm{BZ}^2}{c_\mathrm{TA} } \Theta (\omega - c_\mathrm{TA} q_\mathrm{BZ}) \Theta (c_\mathrm{LA} q_\mathrm{BZ} - \omega) \\
\nonumber
&+ \frac{-2 (c_\mathrm{LA}^3 + c_\mathrm{TA}^3) q_\mathrm{BZ}^3 + 3 (c_\mathrm{LA}^2 + c_\mathrm{TA}^2 ) q_\mathrm{BZ}^2 \omega -  \omega^3 }{ c_\mathrm{TA}^3} \\
&\quad \times \Theta (\omega - c_\mathrm{LA} q_\mathrm{BZ}) \Theta ((c_\mathrm{LA} + c_\mathrm{TA}) q_\mathrm{BZ} - \omega ) \Bigg).
\end{align}
as well as the $\mathrm{LOLA}$ term,
\begin{align}
\label{eq:lola-incoh}
&\mathcal{S}_\mathrm{LOLA} =\frac{2\pi}{\Omega_c} \frac{a^5 \left(q_\mathrm{BZ}^2 q^4 \right)}{768 \pi^5 c_\mathrm{LA}^3 m_p^2 \omega_\mathrm{LO}} (\omega - \omega_\mathrm{LO} ) \nonumber \\
 &\times \Theta(\omega - \omega_\mathrm{LO}) \Theta((c_\mathrm{LA} q_\mathrm{BZ} + \omega_\mathrm{LO}) - \omega).
\end{align}
Again we may find $\mathcal{S}_\mathrm{LOTA}$, $\mathcal{S}_\mathrm{TOLA}$, and $\mathcal{S}_\mathrm{TOTA}$ by relabelings and inserting relevant factors of two for polarizations. Note that, since the incoherent approximation does not recover the $q^6$ scaling resulting from interference, we have written the structure factor here using $q_\mathrm{BZ} = 2\pi / a$ to make the comparison more explicit. At lowest order in $x$ and for $A_1 \approx A_2$, such a comparison of~\eqref{eq:coh-optical-acoustic} and ~\eqref{eq:lola-incoh} shows a relative factor of $40/\pi^3 \approx 1$ for the LOLA channel. Lastly, for the remaining optical-optical channels we find
\begin{align}
\label{eq:optical-optical-incoherent}
\nonumber
\mathcal{S}_\mathrm{LOLO} &= \frac{2\pi}{\Omega_c} \frac{q^4 }{144 m_p^2 \omega_\mathrm{LO}^2}  \delta(\omega - 2 \omega_\mathrm{LO} ) \\
\nonumber
\mathcal{S}_\mathrm{LOTO} &= \frac{2\pi}{\Omega_c}   \frac{ q^4 }{36 m_p^2 \omega_\mathrm{LO} \omega_\mathrm{TO} } \delta(\omega - (\omega_\mathrm{LO} + \omega_\mathrm{TO}) )\\
\mathcal{S}_\mathrm{TOTO} &= \frac{2\pi}{\Omega_c}  \frac{ q^4 }{36 m_p^2 \omega_\mathrm{TO}^2} \delta(\omega - 2 \omega_\mathrm{TO} ).
\end{align}

A comparison now of~\eqref{eq:coh-optical-optical} and~\eqref{eq:optical-optical-incoherent} shows the incoherent approximation gives a smaller structure factor by factors of $2 \pi/5$ -- $6\pi/5 \approx 2$ -- $4$.


\section{Impulse approximation}
\label{app:impulse}

In this section we discuss how to obtain the impulse approximation form of the structure factor, \eqref{eq:impulseapprox} in Sec.~\ref{sec:impulse_approx}. To achieve this we must approximate the $t$ integral in \eqref{eqn:incoh} for large $q$. The expression in \eqref{eqn:incoh} can be written as
\begin{equation}
    \label{eqn:incohapp}
    \mathcal{C}_{\bfl d} = \frac{1}{V} e^{-2 W_d(\mathbf{q})} \! \int\displaylimits_{-\infty}^{\infty} \! \! dt\,  e^{ f(t)}.
\end{equation}
with
\begin{align}
\label{eq:re_im_exponent}
\nonumber
& \mathrm{Re}[f(t)] \equiv \frac{q^2}{2m_d} \int d \omega' \frac{ D_d(\omega')}{\omega'} \cos (\omega' t ) 
\\
& \mathrm{Im}[f(t)] \equiv \frac{q^2}{2m_d} \int d \omega' \frac{ D_d(\omega')}{\omega'} \sin (\omega' t) - \omega t.
\end{align}
From this, we see there is a global maximum in the real part and a global minimum in the modulus of the imaginary part at $t=0$. This allows us to perform a steepest-descent expansion about $t=0$, giving
\begin{equation}
\label{eqn:cld_impulse}
    \mathcal{C}_{\bfl d} \approx \frac{1}{V} \int\displaylimits_{-\infty}^{\infty} \! \! dt \,  e^{i t ( \frac{q^2}{2m_d} - \omega ) - \frac{t^2}{2} \frac{q^2 \bar \omega_d}{2m_d} },
\end{equation}
where again $\bar \omega_d = \int d\omega' \omega' D_d(\omega')$. Note that the leading term in the expansion about $t=0$ cancelled the Debye Waller factor, assuming the form given in \eqref{eq:DWapprox}.
Evaluating the above gives
\begin{equation}
    \mathcal{C}_{\bfl d} \approx \frac{1}{V} \sqrt{\frac{2 \pi}{\Delta_d^2}} e^{- \frac{ \big(\omega - \frac{q^2}{2m_d} \big)^2}{2 \Delta_d^2}},
\end{equation}
which is the impulse approximation result.

In obtaining this form, we have assumed that any other local maxima in $t$ gives a subdominant contribution to the $t=0$ maximum. In particular, aside from the $t=0$ point, which is a global maximum in ${\rm Re}[f(t)]$, there are local maxima in the real part which will generally be near integer multiples of $ 2\pi/\bar{\omega}_d$. The leading order contribution from each additional maxima $t_{\mathrm{max}}$ is given by evaluating the real part in the exponential at the location of the maxima.

This must necessarily be smaller than the $t=0$ contribution since the following inequality is always satisfied
\begin{equation}
    \int d \omega' \frac{D_d(\omega')}{\omega'} \cos (\omega' t_{\rm max} ) <  \int d \omega' \frac{D_d(\omega')}{\omega'}.
\end{equation}
Since $t_{\rm max} \sim 2\pi/\bar \omega_d$, the left hand side will be suppressed by an $O(1)$ amount due to presence of the $\cos (\omega' t_{\rm max} )$. Then, the contribution from the local maxima will be exponentially suppressed:
\begin{equation}
    \label{eq:impulse_maxima}
    e^{\frac{q^2}{2m_d} \int d \omega' \frac{D_d(\omega')}{\omega'} \cos (\omega' t_{\rm max})} \ll e^{\frac{q^2}{2m_d} \int d \omega' \frac{D_d(\omega')}{\omega'}}
\end{equation}
as long as the following condition is satisfied
\begin{equation}
    \label{eq:impulse_approx_condition}
    \frac{q^2}{2m_d} \gg \frac{1}{\int d \omega' \frac{ D_d(\omega')}{\omega'} } \sim \bar \omega_d.
\end{equation}
Here we have taken $\int d \omega' \frac{ D_d(\omega')}{\omega'} \sim 1/\bar \omega_d$ as a typical scale for this integral, although it will differ by an $O(1)$ factor. Therefore, as long as the free nuclear recoil energy $\omega = q^2/(2m_d)$ is well above the typical phonon energy $\bar \omega_d$ for a scattering off of atom $d$, the $t=0$ maximum is dominant and the impulse approximation should be accurate.

In the regime where $q^2/2m_d$ is comparable to $\bar{\omega}_d$, the contributions from the additional maxima in $t$ can become important.  Nevertheless, the impulse approximation is still accurate at large $\omega$ even in this case because of cancellations from the rapidly changing phase in ${\rm Im}[f(t)]$. When $\omega \gg \bar{\omega}_d$, then ${\rm Im}[f(t)] \approx - \omega t$ for $t$ around $t_{\rm max} \sim 2\pi/\bar{\omega}_d$. This implies large oscillations of $f(t)$ around $t_{\rm max}$, which suppresses the contribution from these local maxima. On the other hand, if $\omega \lesssim \bar \omega_d$, there may be large corrections to the impulse approximation due to these additional maxima.

These effects were shown in Fig.~\ref{fig:impulse-structures_top} when comparing the multiphonon expansion result to the impulse approximation. The middle panel showed the result if $q = \sqrt{2 m_d \bar \omega_d}$, in the $m_{\text{Ga}}\approx m_{\text{As}}$ approximation. For $\omega \gtrsim \bar{\omega}_d$ the structure factor falls smoothly and can be reasonably captured by the impulse approximation, while for $\omega \lesssim \bar{\omega}_d \approx 22$ meV or at the optical phonon energies $31$ and $33$ meV there are sharp peaks in the multiphonon response that are not captured by the impulse approximation. For $q = 2\sqrt{2 m_d \bar \omega_d}$ the many multiphonon peaks merge and add up to a shape similar to the impulse approximation over the whole $\omega$ range. Practically, for our calculations, we use the impulse approximation for the structure factor at $q > 2 \sqrt{2 m_d \bar{\omega}_d}$. Though the approximation has small differences with the exact result when $q \sim 2 \sqrt{2 m_d \bar{\omega}_d}$, integrating over the allowed phase space for the rate largely washes out these differences.


\section{Implementation in \texttt{DarkELF}}
\label{app:results_darkelf}

\begin{table*}[bt]
\begin{tabular}{p{0.35\textwidth}p{0.40\textwidth}p{0.18\textwidth}}
\multicolumn{3}{c}{DM-multiphonon scattering}\\\hline
function&description& available for\\\hline
\verb+dRdomega_multiphonons_no_single(omega)+&Differential rate $dR/d\omega$ in 1/kg/yr/eV&all except $\mathrm{SiO}_2$, $\mathrm{Al}_2\mathrm{O}_3$ \\
& excluding long-wavelength single phonons & \\
\verb+R_multiphonons_no_single(omega)+&Total phonon  rate in 1/kg-yr&all except  $\mathrm{SiO}_2$, $\mathrm{Al}_2\mathrm{O}_3$ \\
& excluding long-wavelength single phonons & \\
\verb+sigma_multiphonons(omega)+&Nucleon cross section to produce 3 events/kg-yr &all except  $\mathrm{SiO}_2$, $\mathrm{Al}_2\mathrm{O}_3$ \\\hline
\end{tabular}
\caption{List of public functions in \code\ related to multiphonon excitations from DM scattering. Only mandatory arguments are shown; for optional arguments and flags, see text and the documentation in repository. Some functions are only available for select materials, as indicated in the righthand column. \label{tab:darkelf} }
\end{table*}

In the main text, we presented the formulas in the manner which is most clear from the point of view of the various approximations and their regimes of validity. These formulas were not always suitable however for an efficient numerical implementation, which we address in this section. We also provide details on their implementation in the \texttt{DarkELF} package~\cite{Knapen:2021bwg}.

\begin{figure}
\centering
\includegraphics[width=0.9\linewidth]{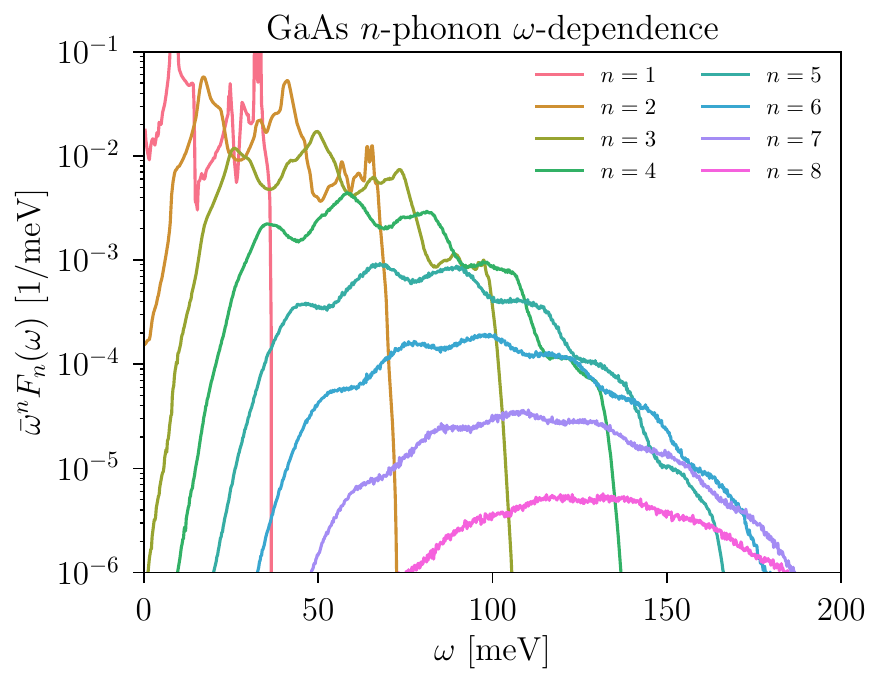} 
\caption{\label{fig:n-phonon-omega-dependence} Here we have plotted $\bar{\omega}^n F_n(\omega)$, where $F_n(\omega)$  is the $\omega$-dependent part of the structure factor in the incoherent approximation and given explicitly in \eqref{eq:Fn_omega}. At fixed $q$, the structure factor decreases quickly with increasing $\omega$.}
\end{figure}

In the main text we gave the rate in the isotropic limit, \eqref{eq:rate-isotropic}. In order to calculate the rate for any mediator and to obtain the differential rate $dR/d\omega$, it is convenient to perform the $v$-integral first and rewrite the rate as:
\begin{equation}
    \label{eq:total_rate}
    R =  \frac{1}{4\pi \rho_T} \frac{\rho_\chi}{m_\chi}  \frac{\sigma_p}{\mu_\chi^2}   \int\displaylimits_{\omega_\mathrm{th}}^{\omega_+} \! d\omega \,  \int\displaylimits_{q_-}^{q_+} \! dq \,   q \, |\Tilde{F}(q)|^2 S(q, \omega) \, \eta(v_{\rm min}(q,\omega))
\end{equation}

where now the integration limits are given by  \begin{align}
    q_\pm &= m_\chi \big(v_\mathrm{max} \pm \sqrt{v_\mathrm{max}^2 - \frac{2 \omega}{m_\chi}} \big) \\
    \omega_+  &= \frac{1}{2} m_\chi v_{\rm max}^2 
\end{align}
with $v_{\rm max} = v_{\rm esc} + v_e$ the maximum DM speed in the lab frame. The $\eta$ function is given by
\begin{equation}
    \label{eq:eta_function}
    \eta (v_\mathrm{min}) = \int d^3 \mathbf{v} \frac{f(\mathbf{v}) }{v} \Theta (v - v_\mathrm{min})
\end{equation}
with  $v_\mathrm{min}(q, \omega) = \frac{q}{2m_\chi} + \frac{\omega}{q}.$

To evaluate the rate using incoherent approximation, we provide look-up tables for the structure factor.
At each $n$ for the sum in \eqref{eqn:n-incoherentstructure}, the $q$ and $\omega$ parts of the integral are separable, so we can capture the  $\omega$-dependent part with the family of functions
\begin{equation}
\label{eq:Fn_omega}
    F_{n,d}(\omega) \equiv \frac{1}{n!} \left(\prod\displaylimits_{i=1}^n \int d \omega_{i} \frac{D_d(\omega_i)}{\omega_i} \right) \delta \left(\sum_i \omega_i - \omega \right),
\end{equation}and calculate the rate in terms of functions $F_{n,d}$. These functions are simple to calculate numerically up to $n \leq 10$, which we have tabulated and provided in \texttt{DarkELF} as look-up tables to speed up the calculation. The combination $\bar{\omega}^n F_n (\omega)$ is shown in Fig.~\ref{fig:n-phonon-omega-dependence} for GaAs in the $m_\mathrm{Ga} \approx m_\mathrm{As}$ approximation. For increasingly high $n$, the $F_{n,d}$ become increasingly smooth.

We have added several additional functions to  \texttt{DarkELF} for the differential and integrated rate calculations from the single phonon to the nuclear recoil regime. Tab.~\ref{tab:darkelf} describes some of the new relevant functions. These functions currently work for materials with up to two atoms per unit cell. We have included the necessary data files for the multiphonon calculation for GaN, Al, ZnS, GaAs, Si, and Ge from a combination of DFT and experimental sources. We also allow the user to input their own calculations or extractions of the (partial) density of states, as well as momentum-dependent dark matter-nucleon couplings. Before calculating multiphonon scattering rates in \texttt{DarkELF}, it is necessary to tabulate the auxiliary function \eqref{eq:Fn_omega} for each atom. This is done using the \texttt{DarkELF} function \emph{create\_Fn\_omega}. This step is the most time consuming part of the calculation, so we provide these pre-tabulated for the aforementioned materials. For calculations with a user-supplied (partial) density of states, these tables must first be updated by running \emph{create\_Fn\_omega}. \texttt{DarkELF} will save these new look-up tables for future computations, such that this step only need to be performed once. Next we describe the functions that return important results. All of the following straightforwardly apply equations (\ref{eq:total_rate}-\ref{eq:eta_function}).

\emph{R\_single\_phonon}: This function takes the energy threshold and DM-nucleon cross sections and outputs the rate in the long-wavelength single phonon regime using the analytic functions (\ref{eq:single-ph-aco-analytic}-\ref{eq:single-ph-opt-analytic}).

\emph{R\_multiphonons\_no\_single}: This function takes the energy threshold and DM-nucleon cross section as inputs and calculates the total integrated rate, excluding the single phonon processes at long wavelengths $q < q_\mathrm{BZ}$. In other words, this calculation includes only the purple (multiphonon expansion) and red (impulse approximation) phase space regions in Fig.~\ref{fig:illustrative_plot}.

\emph{sigma\_multiphonons}: This takes the energy threshold as input and returns the necessary DM-nucleon cross section to produce three events per kg-year for any number of phonons. In order to return this cross section, this function first calculates the total rate by summing the outputs of \emph{R\_single\_phonon} and \emph{R\_multiphonons\_no\_single}, so it includes the entire calculation scheme depicted in Fig.~\ref{fig:illustrative_plot}.

\emph{\_dR\_domega\_multiphonons\_no\_single}: This function takes the energy transfer $\omega$ and DM-nucleon cross section and returns the differential rate $\frac{dR}{d\omega}$ at that energy excluding single phonons in the long wavelength regime. This comes from equation \eqref{eq:total_rate} without evaluating the $\omega$ integral. We exclude the single coherent phonon here since the long-wavelength approximation has delta functions in energy in the differential rate.

\section{Additional results\label{app:extraresults}}

\begin{figure}
\includegraphics[width=0.98\linewidth]{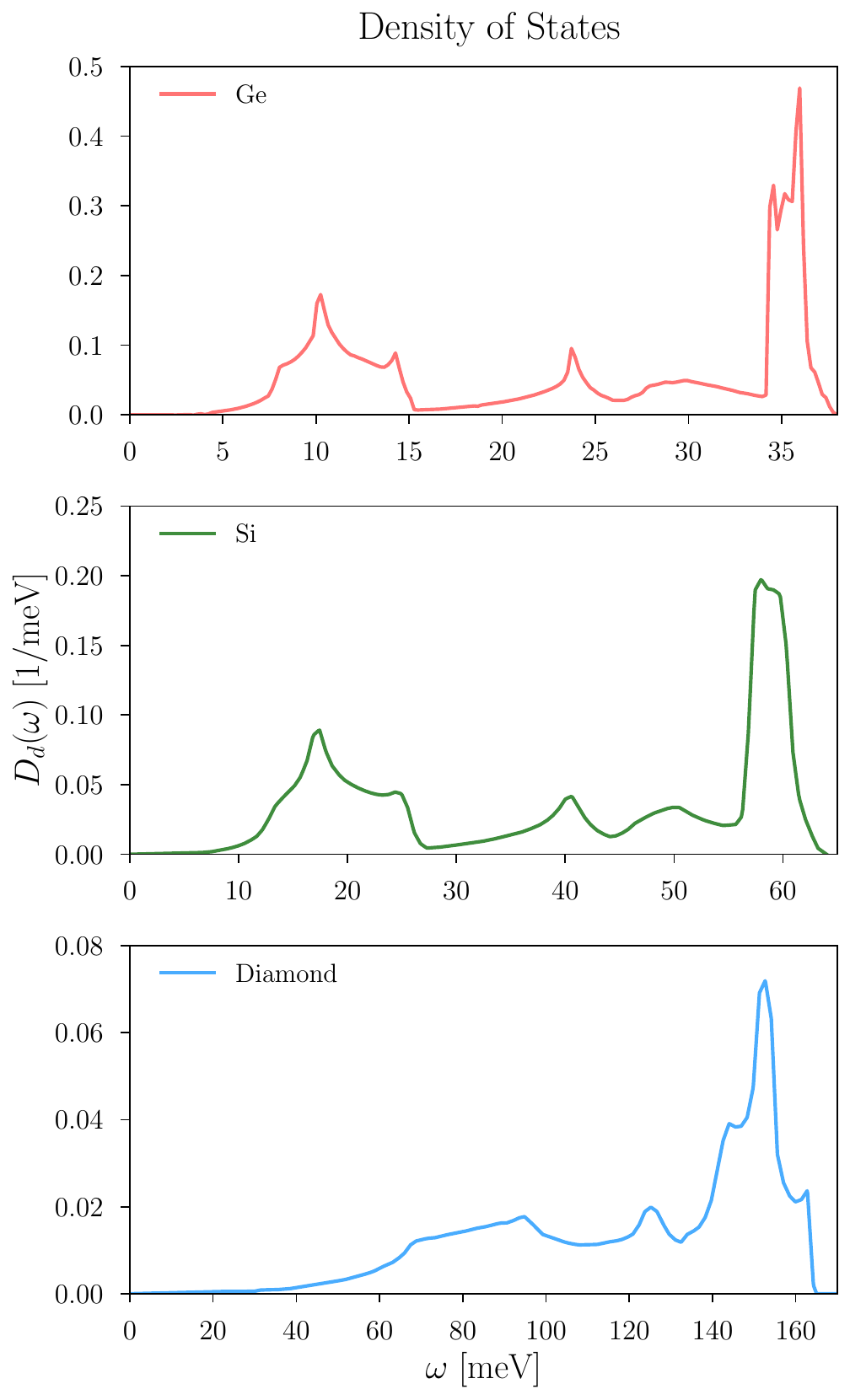}
\caption{Densities of states for germanium, silicon, and diamond~\cite{Jain2013}. 
\label{fig:densityofstates_materials}
}
\end{figure}

\begin{figure*}
\centering
\includegraphics[width=0.97\linewidth]{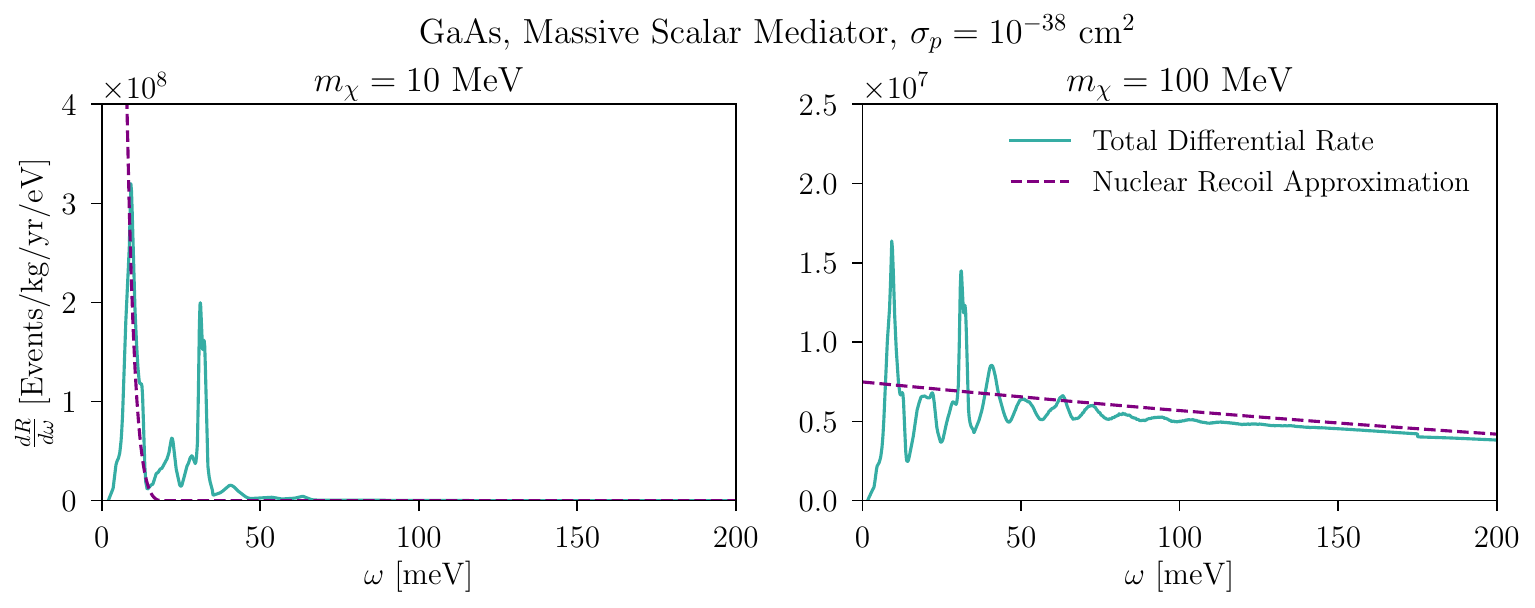} 
\includegraphics[width=0.97\linewidth]{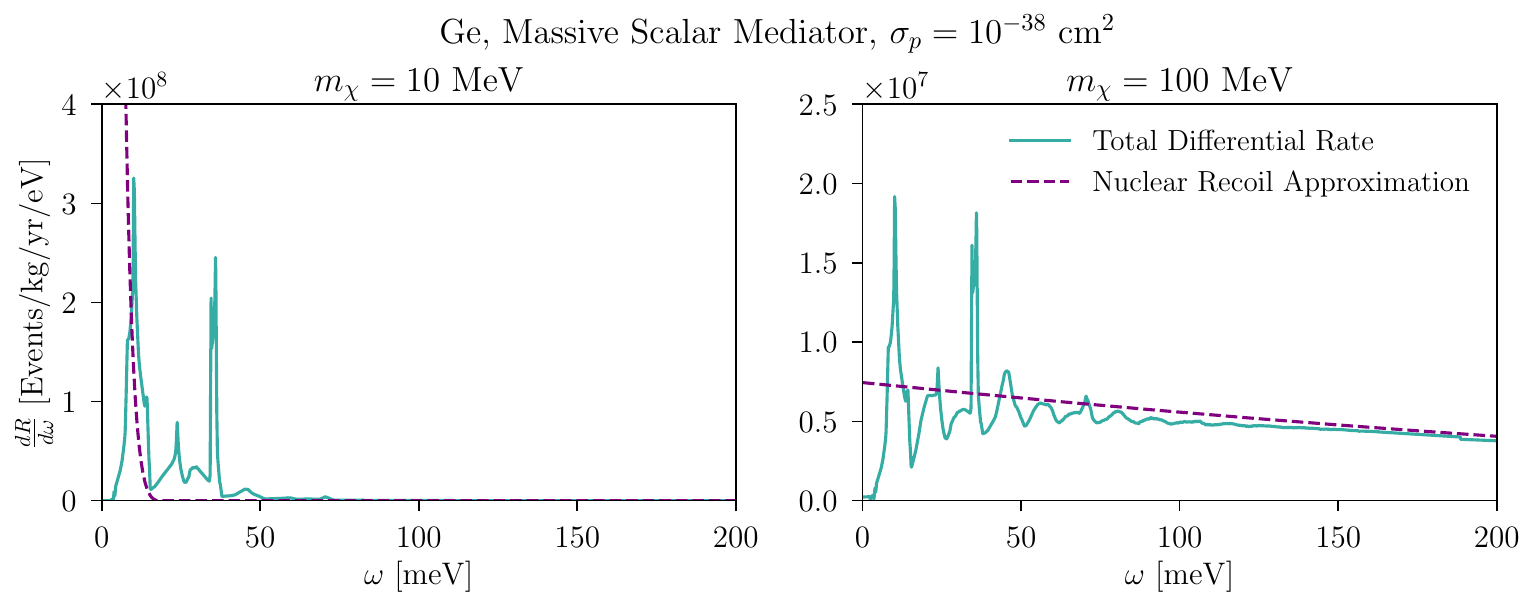} 
\includegraphics[width=0.97\linewidth]{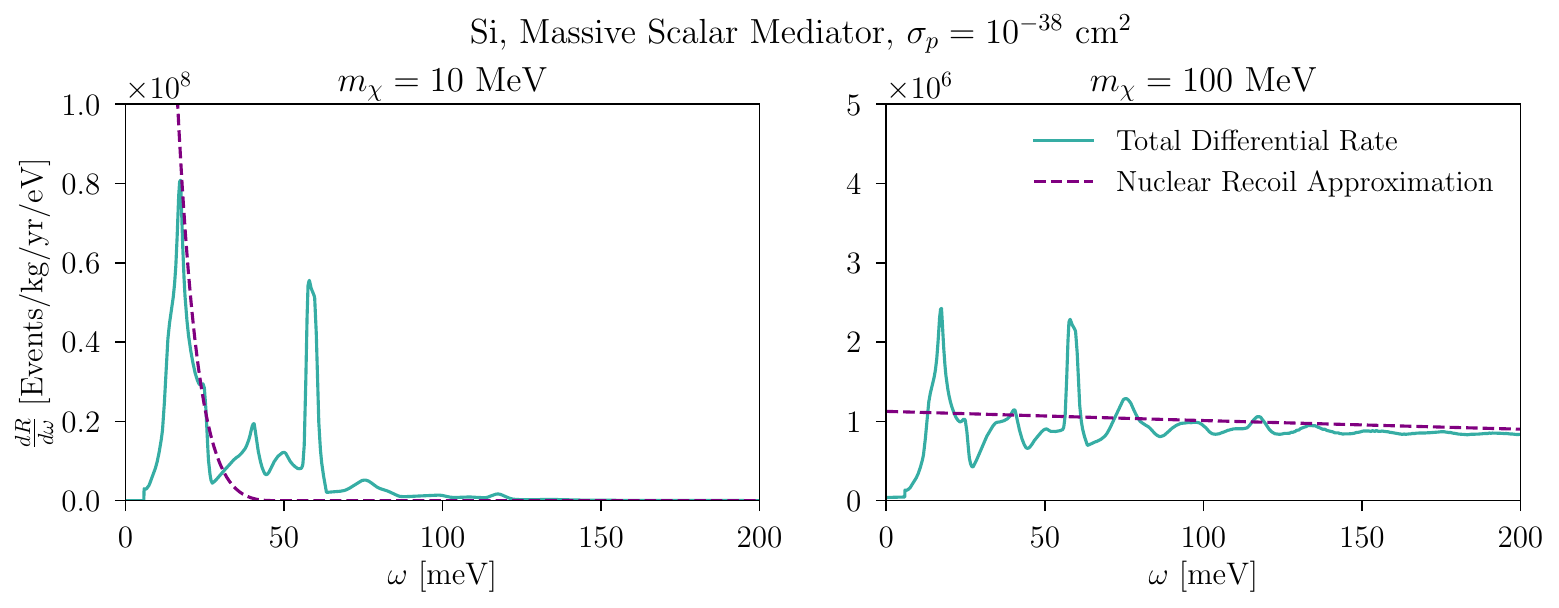} 
\caption{  Differential rate for various materials and a massive scalar mediator, compared with the nuclear recoil approximation. The single phonon contribution from the long wavelength regime is not shown, since it gives a delta function contribution.
\label{fig:differential_rate}}
\end{figure*}

\begin{figure*}
\centering
\includegraphics[width=0.47\linewidth]{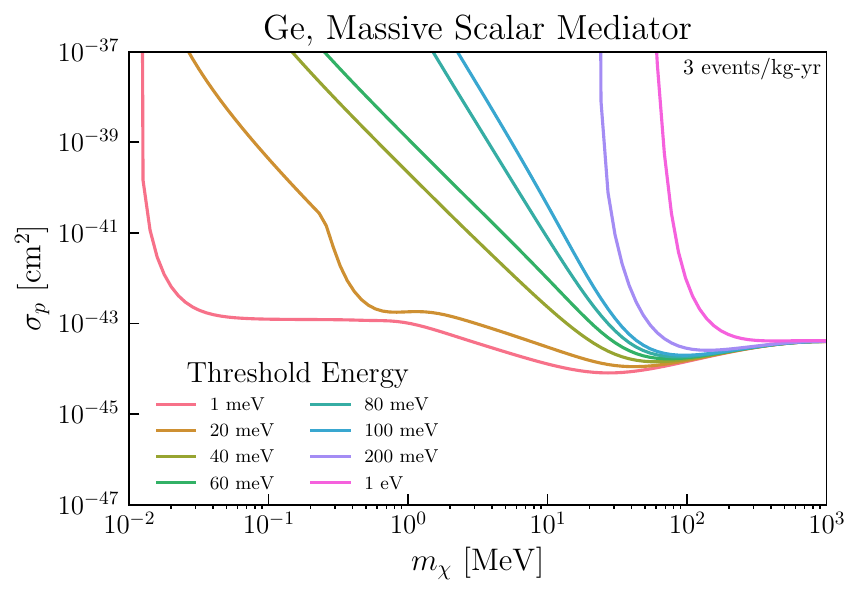}
\includegraphics[width=0.47\linewidth]{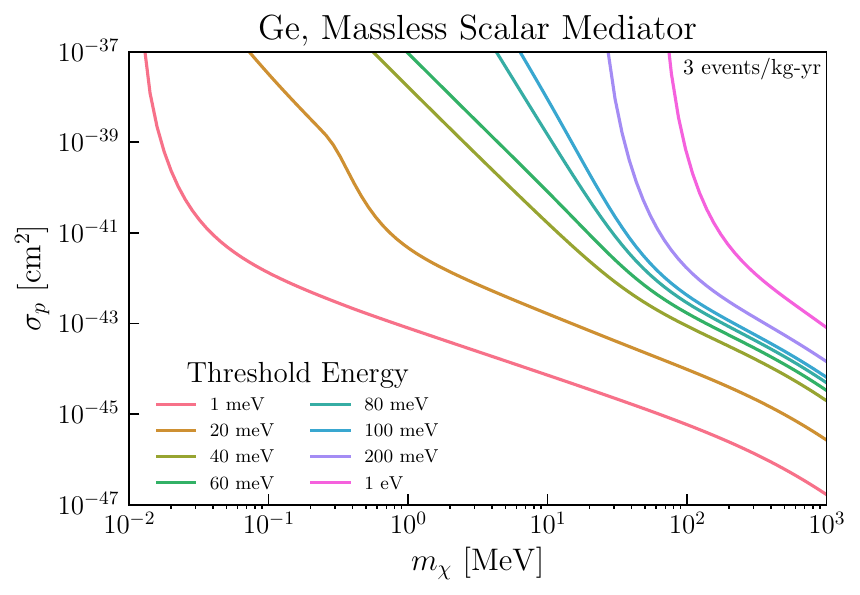}
\includegraphics[width=0.47\linewidth]{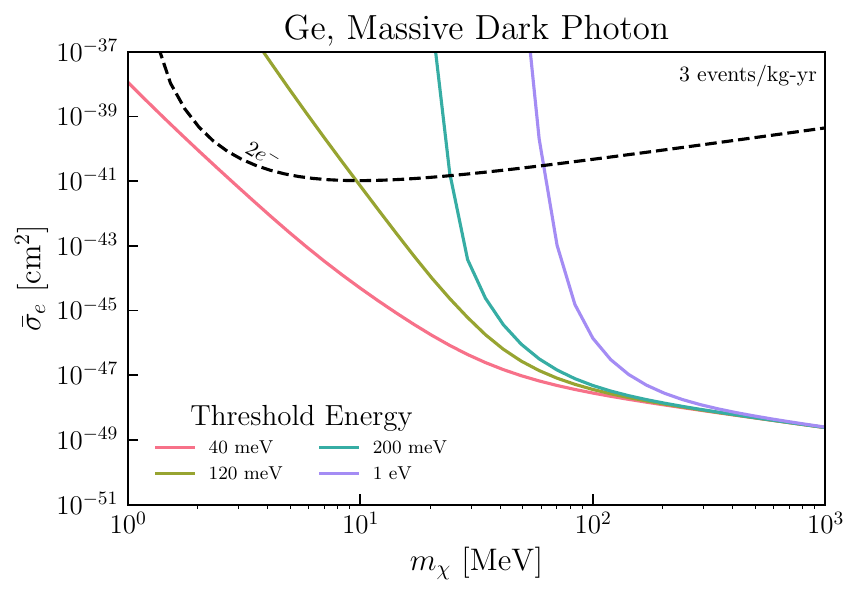} 
\includegraphics[width=0.47\linewidth]{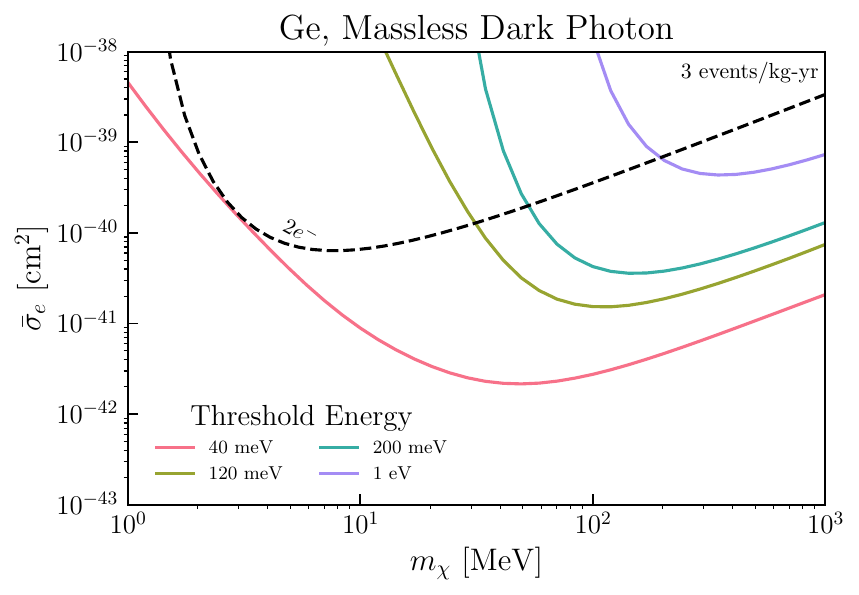}

\caption{ Cross section plots for a rate of 3 events/kg-year exposure for different thresholds in Ge.\label{fig:ge_reaches}}
\end{figure*}

\begin{figure*}
\centering
\includegraphics[width=0.47\linewidth]{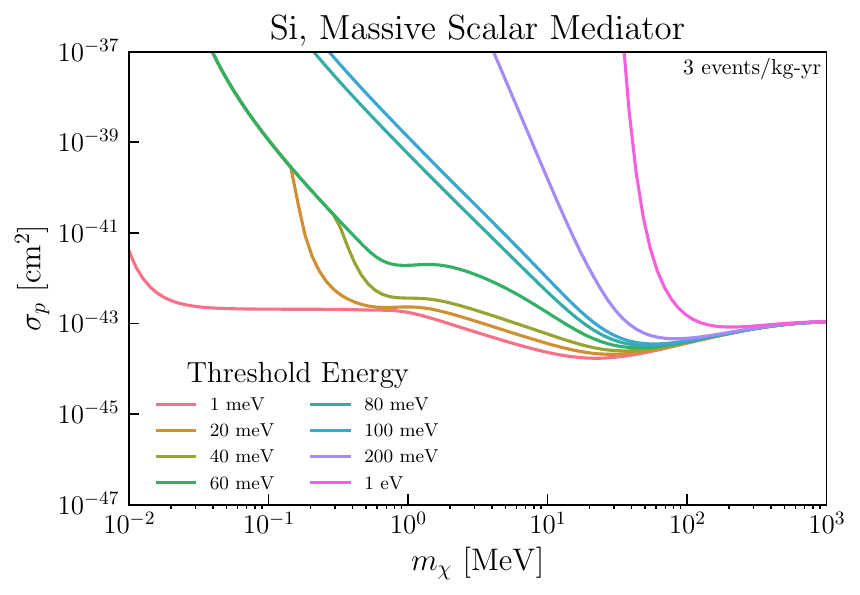} 
\includegraphics[width=0.47\linewidth]{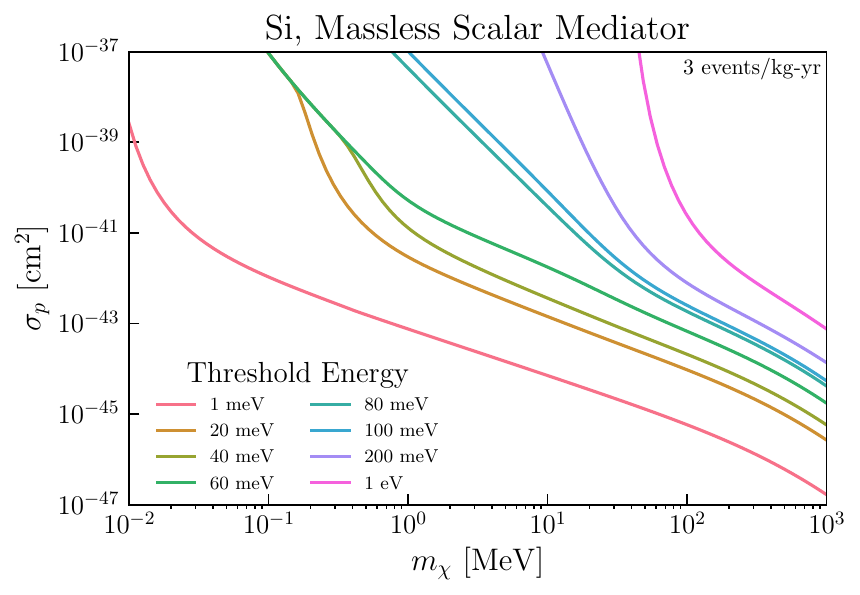} 
\includegraphics[width=0.47\linewidth]{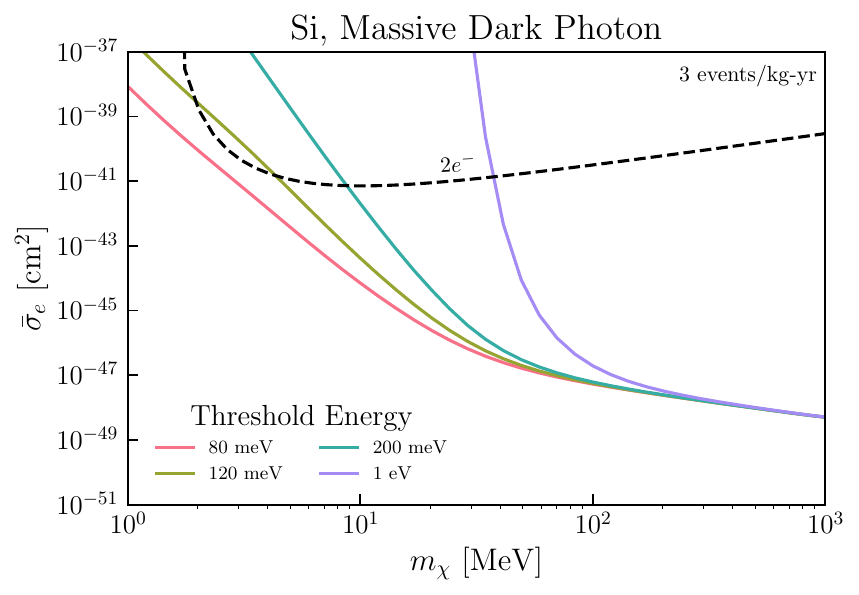} 
\includegraphics[width=0.47\linewidth]{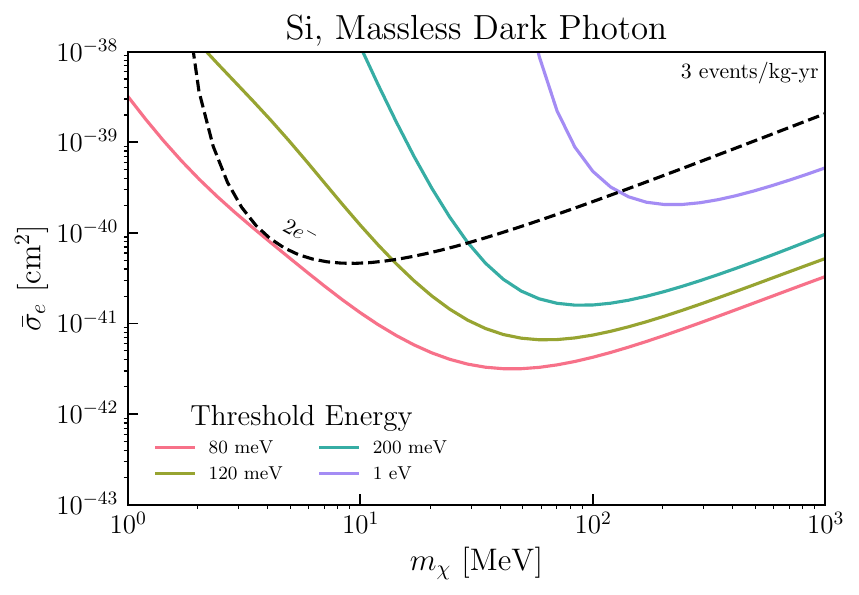} 

\caption{Cross section plots for a rate of 3 events/kg-year exposure for different thresholds in Si.
\label{fig:si_reaches}}
\end{figure*}

\begin{figure*}
\centering

\includegraphics[width=0.47\linewidth]{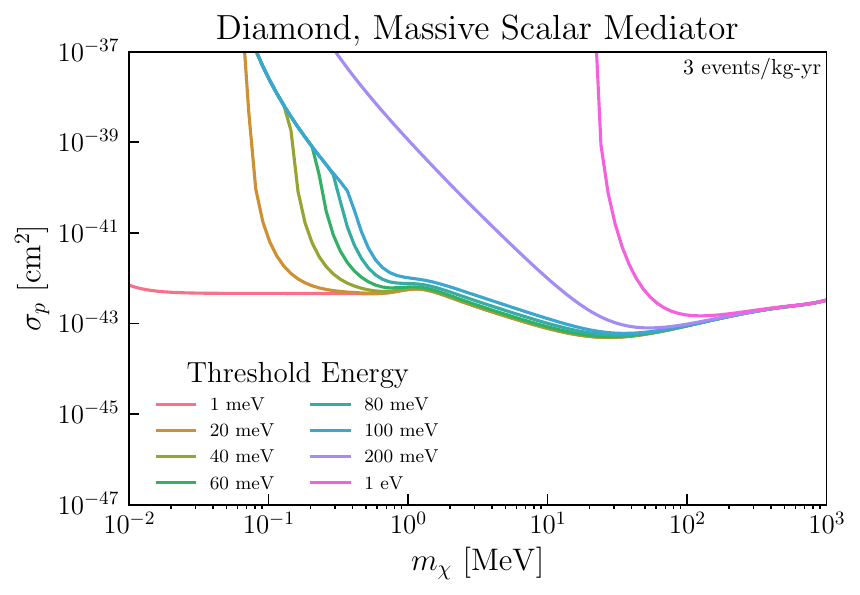} 
\includegraphics[width=0.47\linewidth]{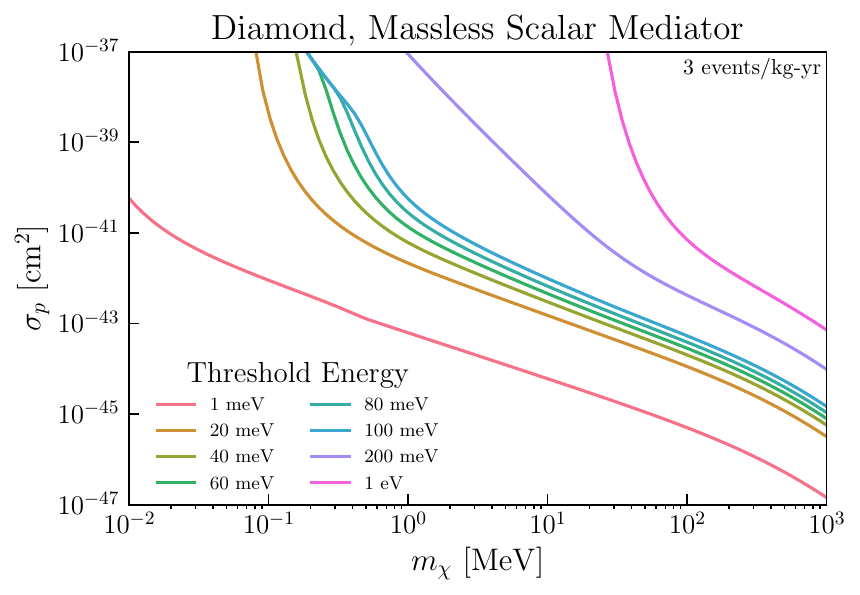}
\includegraphics[width=0.47\linewidth]{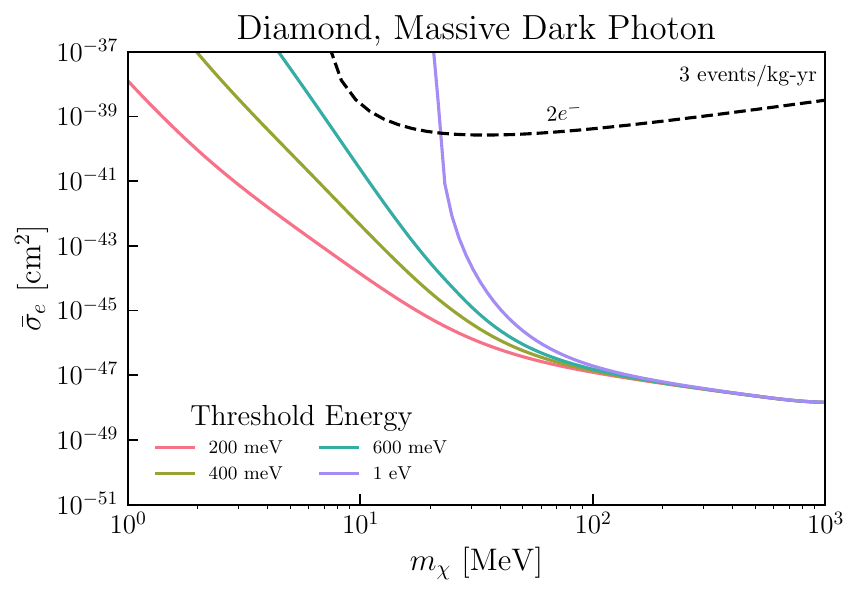} 
\includegraphics[width=0.47\linewidth]{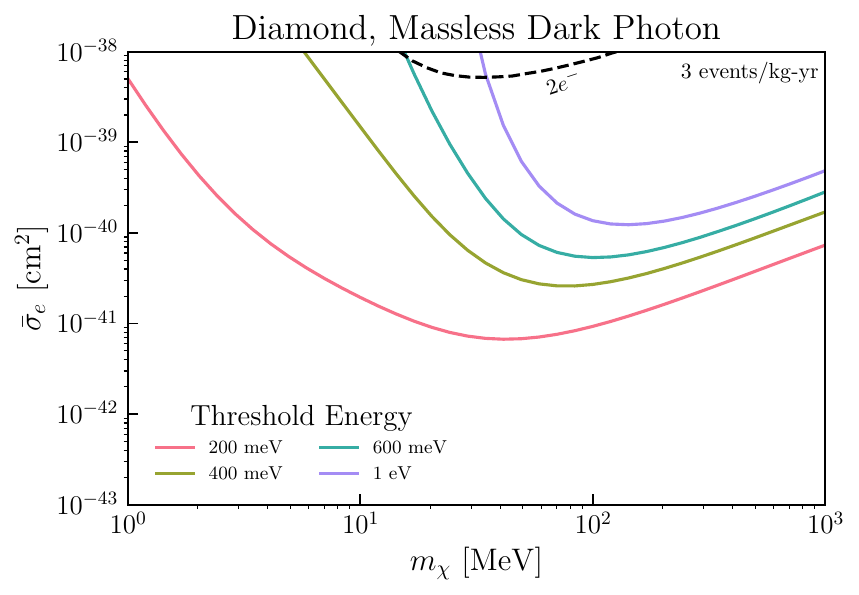}
\caption{ Cross section plots for a rate of 3 events/kg-year exposure for different thresholds in diamond.
\label{fig:diamond_reaches}}
\end{figure*}

Here, we provide additional results for Ge, Si,  and diamond. Concretely, Fig.~\ref{fig:densityofstates_materials} shows the density of states for these three materials, as extracted from~\cite{Jain2013}. Fig.~\ref{fig:differential_rate} shows the differential scattering rate via a massive scalar mediator for two example DM masses in GaAs, Ge and Si targets. Finally, Figs.~\ref{fig:ge_reaches},~\ref{fig:si_reaches}, and~\ref{fig:diamond_reaches} are the cross section plots corresponding to an integrated rate of 3 events/kg-year for Ge, Si, and diamond, respectively. The electron recoil cross sections shown (dashed black lines) are based on calculations in \cite{Knapen:2021run} for Ge, Si and in \cite{Liang:2020ryg} for diamond.

\clearpage

\bibliographystyle{apsrev4-1}
\bibliography{nphonon}

\end{document}